# Multiple scattering theory for polycrystalline materials with strong grain anisotropy: theoretical fundamentals and applications

Huijing He[†]

Department of Mechanical and Materials Engineering

University of Nebraska-Lincoln, Lincoln, Nebraska 68588, USA

[†] he.hui.jing@hotmail.com

## ABSTRACT

This work is a natural extension of the author's previous work: "Multiple scattering theory for heterogeneous elastic continua with strong property fluctuation: theoretical fundamentals and applications" (arXiv:1706.09137 [physics.geo-ph]), which established the foundation for developing multiple scattering model for heterogeneous elastic continua with either weak or strong fluctuations in mass density and elastic stiffness. Polycrystalline material is another type of heterogeneous materials that widely exists in nature and extensively used in industry. In this work, the corresponding multiple scattering theory for polycrystalline materials with randomly oriented anisotropic crystallites is developed. To validate the theory, the theoretical results for a series of materials such as OFHC copper, 304 stainless steel, and Inconel 600 are compared to experimental measurements and the numerical results obtained using finite element simulations. Detailed analysis shows that the new theory is capable of predicting the dispersion and attenuation of polycrystals with satisfactory accuracy. The results also show the new model can give an estimate on the average grain size with a relative error equal to or less than ten percent. As applications in ultrasonic nondestructive evaluation, we calculated the dispersion and attenuation coefficient of one of the most important polycrystalline materials in aeronautics engineering: high-temperature titanium alloys. The effects of grain symmetry, grain size, and alloying elements on the dispersion and attenuation behaviors are examined. Key information is obtained which has significant implications for quantitatively evaluating the average grain size, monitoring the phase transition, and even estimating gradual change in chemical composition of titanium components in gas turbine engines. For applications in seismology, the velocities and Q-factors for both hexagonal and cubic polycrystalline iron models for the Earth's uppermost inner core are obtained in the whole frequency range. Using the realistic material parameters of iron under the high temperature and high-pressure conditions calculated from ab initio simulations, the numerical results show that the Q-factors range from 0.001 to 0.05, which shows good agreement with that inferred from real seismic data. The new model predicts the velocity of longitudinal waves varies between ± 1% to ± 5 % relative to the Voight average velocity, while the velocity of transverse waves varies from ± 10% to ± 20%, which gives promising explanation to the abnormally slow transverse velocity observed in practical measurements. The numerical results support the conjecture that the Earth's uppermost inner core is a solid polycrystalline medium. The comprehensive numerical examples show the new model is capable of capturing the most important scattering features of both ultrasonic and seismic waves with satisfactory accuracy. This work provides a universal, quantitative model for characterization of a large variety of polycrystalline materials. It also can be extended to incorporate more complicated microstructures, including ellipsoidal grains with or without textures, and even multi-phase polycrystalline materials. The new model demonstrates great potential of applications in ultrasonic nondestructive evaluation and inspection of aerospace and aeronautic structures. It also provides a theoretical framework for quantitative seismic data explanation and inversion for the material composition and structural formations of the Earth's inner core.

## I. INTRODUCTION

Polycrystalline materials constitute a class of most important heterogeneous materials that both widely exist in nature and extensively used in industry. The majority of metallic minerals on the Earth exist in the form of polycrystalline compounds. In modern electronic industry, sintered and poled ceramic like Lead Zirconate Titanate (PZT) is an important type of piezoelectric polycrystalline materials, which has found broad applications in manufacturing smart materials and structures. Due to the outstanding mechanical, thermal and chemical properties, polycrystalline high-temperature alloys have become the key material in modern aeronautics and aerospace industries for manufacturing high performance gas turbine engine of commercial and military aircrafts [1-3]. As the performance of modern jet fighters and commercial aircraft continues improving, the requirements for the jet engine become more and more demanding. For instance, the supersonic stealth fighter F-22 reaches a cruising speed up to Mach 2.25, which requires the thrust-to-weight ratio of the turbine jet engine exceed 10. To meet these stringent requirements, the turbine engine must be capable of working in high-temperature and high-pressure environment for a long time. For example, the working temperature of the turbine engine F119 manufactured by Pratt & Whitney company can now achieve 600 °C, meanwhile, its service life still can be maintained for more than 10000 hrs. The key point to achieve these technological requirements lies in the development of reliable high-temperature materials. Up to now, jet engine manufactures have developed a large quantity of high-temperature alloys and superalloys, most of which are titanium-based or nickel-based, such as Ti64 (Ti-6Al-4V), Ti6242 (Ti-6Al-2Sn-4Zr-2Mo), and most recently, the nonburning titanium alloy named alloy-C (Ti-35V-15Cr-0.05C). Metallurgical studies reveal that the most important mechanical and thermal properties like high strength, outstanding fatigue resistance, and extraordinary resistance to high temperature and high pressure are all closely related to the microstructures, including the grain size, grain shape, preferred orientations of the grain crystallographic axes (textures), and their phase composition [1-2, 4]. In-situ monitoring the formation of microstructures during processing helps engineers get a better understanding of the formation mechanism and evolution of microstructures, which is of significant importance for further improving the processing procedure, adjusting the operating parameters to ensure that the desired microstructures are achieved. Therefore, studying advanced technologies for the nondestructive evaluation and characterization of key materials is of vital importance for the development of modern jet engines. Moreover, nondestructive characterization of microstructures also plays a critical role in jet engine maintenance. Currently the integrally bladed disks (IBDs), in which the blades and the compressor disk are manufactured into a single-piece construction, are widely used in the fan and compressor sections of state-of-the-art military turbine engines due to their desirable features such as fewer number of parts and outstanding resistance to fatigue cracks. Despite their exceptional performance, they also have several drawbacks, such as large size, complex structures, and extremely expensive manufacturing cost. More importantly, they are frequently subjected to foreign object damages. Different repair procedures like additive processing for minor damage and welding of patches for significant damage are often used to renovate the damaged IBDs. In order to restore the original microstructures and minimize the mechanical property mismatch with the original parts, postprocessing heat treatment is usually adopted. Consequently, there is a keen need to develop nondestructive inspection (NDI) technologies to assess the integrity and microstructure state of repaired IBD airfoils. Modern detection technologies, like Electron Backscatter Diffraction (EBSD) [5], Atomic Force Microscopy (AFM) [6] etc. all have been applied to the characterization of microstructures and textures in titanium alloys. However, these characterization modalities often require careful surface preparation and sophisticated measurement equipment. Moreover, these two modalities can only detect the near-surface microstructures. Compared to other characterization approaches, ultrasonic technology possesses a number of unique advantages. For instance, ultrasonic NDE provides an economic, nondestructive, and easy-to-operate approach, in which only piezoelectric transducers, and conventional signal generation and processing systems are used. A more attracting capability is that ultrasonic waves can insonify a significant depth (up to tens of centimeters) into the sample. This unique feature enables the possibility of imaging defects and characterizing microstructures deep in the tested structures. Consequently, extensive research efforts have been devoted to developing ultrasonic nondestructive characterization and inspection techniques [7-24].

    Scattering of seismic waves constitutes another research subject of this work. Seismic observations reveal that both longitudinal and transverse waves can propagate in the inner core, which provide strong evidence that the Earth's inner core is composed of solid materials. Under the action of gravitation, the pressure at the Earth's inner core achieves up to 330-360 GPa,

and the temperature raises up to 5700 K, considering the fact that the core has very high density and the iron is one of the major component element of the Earth, most researchers believe the inner core is mainly composed of iron alloys. The seismic waves scattered from the upper 300 km of the inner core exhibit strong attenuation and isotropic scattering characteristics, thus the uppermost inner core is considered to be composed of untextured polycrystals. The inner part of the inner core shows obvious anisotropic scattering behavior, so it is generally modeled as textured polycrystals. Calvet and Margerin [25] proposed a duplex microstructure model for the uppermost inner core. In this model, the upper 300km of the inner core is assumed to be composed of randomly oriented patches of crystallite clusters, where the crystallites in the same cluster share similar crystallographic orientations. As a primary approximation, each patch is regarded as a single macro grain. On the basis of this model, they calculated the velocity and Q-factors of seismic waves and concluded that the polycrystalline model shows great potential to explain observed seismic wave velocity and attenuation.

Polycrystalline alloys are composed of large quantities of single-crystal elements – grains, also called crystallites. Each type of single crystals belongs to a specific class of crystal symmetry, so it always exhibits anisotropic elastic behavior. Each grain in a polycrystalline material possesses a unique crystallographic orientation and thus, the polycrystal contains a large number of grain boundaries. The mismatch of acoustic impedance at the grain boundaries causes multiple scattering of ultrasonic waves during its propagation. As a consequence, the amplitudes of ultrasonic signals are attenuated and the velocity is frequency-dependent. As the ultrasonic pulse propagating in the material, portion of the scattered energy propagates backward and induces observable grain noises. The transmitted signals and backscattered noises both contain rich information about the microstructures of the sample. The technical significance of ultrasonic scattering in polycrystalline materials has been noticed even since the late fifties of the last century. Bhatia [26-27] calculated the intensity attenuation coefficients for dilatational and shear waves based on the weak-property fluctuation assumption and the low frequency approximation. Since then, significant research efforts have been devoted to exploring the possibility of utilizing ultrasonic technology to characterizing microstructures in polycrystalline materials. In 1979, Ranganathan and Dattagupta [28] proposed a scattering model for ultrasound attenuation in polycrystalline materials based on the weak fluctuation assumption and the Chernov theory. The first multiple scattering theory for elastic waves was developed by Karal and Keller [29] in 1964. With the small perturbation assumption, the differential wave operator and its inverse operator are expanded into an infinite series of the small parameter which represents the property fluctuation. Successive iteration of the series expansion into the elastodynamic equations yields a system of integral equations. The infinite series is truncated and the terms containing up to the second order statistics of the random material properties are retained, which finally leads to the Christoffel equation for the coherent waves. The complex wavenumber for coherent waves are obtained. The method developed in this work is called the Keller approximation. Based on this pioneering work, Stanke and Kino [30-31] developed the corresponding multiple scattering model for polycrystalline materials with cubic crystallites. The dispersion and attenuation curves for cubic polycrystals like aluminum and iron are obtained. Comparison with experimental data reveals the model is capable of predicting the average grain size with a relative error less than 20%. This model is named the unified scattering model since it gives a unified theoretical framework for predicting the propagation behavior in the whole frequency range, covering the Raleigh regime, the stochastic regime and the geometric optic regime. It is worth mentioning that only one propagation mode was found in the whole frequency range. This model has been generalized to characterize polycrystals with more complicated microstructures. Amhed and Thompson [32-39] investigated the effects of texture and grain elongation on the scattering attenuation using Stanke-Kino's model. During 1982 to 1988, Hirsekorn [40-43] conducted comprehensive study on the multiple scattering of elastic waves in various polycrystalline materials, including untextured and textured polycrystals, and multiphase polycrystals. A distinct feature in her theory is that the series expansion is performed on the displacement potentials instead on the displacements. In 1990, Weaver [44] developed a multiple scattering model for the mean and mean square response of untextured polycrystals with cubic crystallites in the framework of Dyson's equation and Bethe-Salpeter equation, respectively. The first-order-smoothing approximation (FOSA) and the ladder approximation are adopted to obtain the explicit expression of the mass operator and the intensity operator, respectively. Closed form expressions for the attenuation coefficients are obtained by further introducing the Born approximation. As pointed out in [44], the small perturbation expansion of the wave operator is introduced and the theory is limited to polycrystals with weak grain anisotropy. In parallel to the development of other scattering theories, this model has also been extended to incorporate polycrystals with more complicated microstructures. Turner extended Weaver's model to the case of textured polycrystals [45]. Kube and Turner [46] further derived the attenuation coefficients under the Born approximation for materials with general anisotropic crystallites. The corresponding scattering theory for locally isotropic media is developed by Turner and Anugonda [47]. The analytical expression of the original dispersion equation is extremely complicated, moreover, solving for its numerical solution is even more challenging, so all the numerical results in the abovementioned works are obtained with the Born approximation, which are valid in the frequency regime lower than the geometric region. The accurate solution of the dispersion and attenuation calculated directly from the FOSA Dyson's equation is a longstanding problem. In order to investigate the dispersion and attenuation behavior in the whole frequency range, Calvet and Margerin [48] introduced the so-called spectral method, in which the imaginary part of the ensemble-averaged Green's function is defined as the spectral function and the velocity and attenuation are extracted from this function via least square approach. This method circumvents the challenging problem of solving the complicated dispersion equations. They first discovered that there is one propagation mode at relatively low frequencies and two modes exist at high frequencies, though in the end of the paper they questioned this discovery. Later on, they studied the effects of grain elongation using the same approach [49]. Calvet and Margerin first applied Weaver's model to study the seismic wave attenuation occurred at the Earth's uppermost inner core [25]. One common feature of Weaver's model and Stanke-Kino's model is that both of them use the Voigt average medium as the homogeneous reference material. It is well-known the Voigt average elastic stiffness gives the upper bonds of the homogenized media, so it always overestimates the quasi-static limits of the coherent waves. To remedy this discrepancy, Kube and Turner introduced a self-consistent scheme to calculate the quasi-static limits of the coherent wave velocities, and proposed to use these material properties as the homogeneous reference in Weaver's model [50].

Through the above discussion we see most of current ultrasonic scattering models are based on the weak scattering assumption, which is valid when the grain anisotropy or texture anisotropy is relatively weak. In this work, the author strives to develop a universal multiple scattering theory for polycrystalline materials with general anisotropic crystallites, regardless of whether the grains are weakly anisotropic. Following the new paradigm created in the previous work [51], the renormalized Dyson's equation for polycrystalline materials are derived in Section II, by using the Fourier transform technique, the closed form solution of the coherent wave field in spacetime is obtained. Meanwhile the dispersion equation for the longitudinal and transverse coherent waves are derived. The dispersion and attenuation of polycrystals with different degrees of grain anisotropy are presented in Section III. The accuracy of the new model is validated through comprehensive comparison with Stanke-Kino's model, the experimental data, and numerical simulations. To show its practical applications in ultrasonic nondestructive characterization, the dispersion and attenuation of pure titanium and Titanium alloys are analyzed in Section IV. Particular attentions are paid to the effects of grain size, phase change, and variations in alloying elements on the propagation parameters. As an example of practical application in seismology, the dispersion and Q-factors of hexagonal and cubic iron models of the Earth's inner core are also calculated and possible explanations to a series of longstanding problems concerning the observed velocity and attenuation of seismic waves from inner core are discussed. Section V gives a detailed discussion on how to conduct more accurate numerical and experimental validation of the new model. We also discussed several important extensions of the current model to incorporate more complicated microstructures that are of practical importance. The unique features of the new model and the major conclusions are highlighted in Section VI.

## II. THEORETICAL FUNDAMENTALS

In this section, we present the theoretical fundamentals of the new multiple scattering theory for polycrystalline materials. The development of the new theory consists of the following major steps: 1) Starting from the elastodynamic equations, the displacement and strain Green's functions of the inhomogeneous medium are expressed in terms of the homogeneous reference Green's function and the material property fluctuation; 2) Eliminate the singularity of the Green's tensor appeared in the strain Green's function, and then obtain a set of renormalized integral equation; 3) Derive the renormalized Dyson's equation by applying Feynman's diagram and first-order-smoothing approximation; 4) Solve the system of integral equations using Fourier transform technique, and obtain the dispersion equations. To keep this work in a self-contained manner, and stress the peculiar points that applied for polycrystals specifically, the major steps are detailed as follows.

The time-harmonic wave propagation in a polycrystalline medium is governed by the classic elastodynamic equation [52]:

$$[c_{ijkl}(\mathbf{x})u_{k,l}]_{,j} + \rho(\mathbf{x})\omega^2 u_i = 0, \tag{1}$$

where $\omega$ is the circular frequency, $u_i$ denotes the displacement components, $\rho(\mathbf{x})$ and $c_{ijkl}(\mathbf{x})$ are the mass density and elastic stiffness. The elastic stiffness has the following symmetries:

$$c_{ijkl}(\mathbf{x}) = c_{jikl}(\mathbf{x}) = c_{ijlk}(\mathbf{x}) = c_{klij}(\mathbf{x}). \tag{2}$$

Throughout this work the Cartesian tensor is used. A bold-faced letter represents a vector, tensor or matrix, and italic letters with subscript indices represent tensor components or matrix elements. A comma followed by a coordinate index means taking partial derivative with respect to the corresponding spatial coordinate. The Einstein summation convention, i.e., a repeated index implies summation over that index from 1 to 3, is assumed in this work.

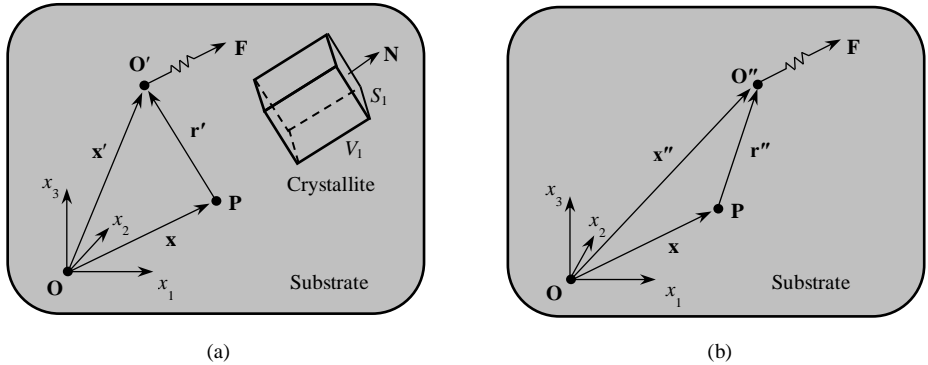

FIG. 1. Definition of the single-inclusion Green's function (a) and the homogeneous Green's function (b).

As the first step in the development of the new theory, we consider an infinite homogeneous elastic medium embedded with a single crystallite, for which the crystallographic axes are oriented arbitrarily, as shown in Fig. 1(a). The quantities pertaining to the homogeneous reference medium and those to the crystallite are discriminated by an attached index "(0)" or "(1)", respectively. In crystallography the orientation of a single crystal relative to a reference, global coordinate system is usually described by the Euler angles. The Euler angles are also frequently adopted to describe the orientation distribution functions of textured polycrystals. Considering its important applications in this work and a series of following researches, the definition of the Euler angles ($\theta$, $\varphi$, $\psi$) and related conventions used in this work are rigorously presented in the appendix.

Green's function is defined as the resulting field excited by a time-harmonic unit concentrated force $\mathbf{F}$ applied at a generic point $\mathbf{O}'$ along $\mathbf{e}_{a'}$, where $\mathbf{e}_{a'}$ represents the coordinate basis of a source-region coordinated system with its origin located at $\mathbf{O}'$. The coordinate basis of the coordinate system $\mathbf{O}'x'y'z'$ by no means needs to be the same as those of the coordinate system $\mathbf{O}xyz$, so at this point we assume they are different from each other. The fields in the substrate and in the crystallite are governed by two sets of different equations, which are given by:

$$c_{ijkl}^{(0)} G_{ka',lj} + \rho^{(0)}\omega^2 G_{ia'} + F a_{ia'}\delta(\mathbf{x}-\mathbf{x}') = 0, \quad \text{in } V_0, \tag{3}$$

$$c^{(1)}_{ijkl}G_{ka',lj} + \rho^{(1)}\omega^2 G_{ia'} + Fa_{ia'}\delta(\mathbf{x}-\mathbf{x}') = 0, \quad \text{in } V_1, \tag{4}$$

where $V_0$ and $V_1$ represent the volumes occupied by the reference medium and the crystallite, respectively, and $a_{ia'}$ denotes the directional cosine of $\mathbf{e}_{a'}$, i.e., $a_{ia'} = \cos(\mathbf{e}_{a'}, \mathbf{e}_i)$.

Suppose the crystallite and the substrate are perfectly bounded, so the stress and displacement are continuous across the boundary of the crystallite, thus we have

$$c^{(0)}_{ijkl}G_{ka',l}(\mathbf{x}\to S^+)N_j = c^{(1)}_{ijkl}G_{ka',l}(\mathbf{x}\to S^-)N_j, \quad G_{ka'}(\mathbf{x}\to S^+) = G_{ka'}(\mathbf{x}\to S^-), \quad \text{on } S. \tag{5}$$

Following the standard procedure as comprehensively presented in [51], we obtain the integral representation of the perturbed field, i.e., the displacement and strain Green's functions of the polycrystalline material composed of an infinite number of randomly oriented crystallites, as follows:

$$\begin{bmatrix} G_{\beta'a'}(\mathbf{x}'',\mathbf{x}') \\ \varepsilon_{\alpha'\beta'a'}(\mathbf{x}'',\mathbf{x}') \end{bmatrix} = \begin{bmatrix} G^0_{\beta'a'}(\mathbf{x}'',\mathbf{x}') \\ \varepsilon^0_{\alpha'\beta'a'}(\mathbf{x}'',\mathbf{x}') \end{bmatrix} + \int_V \begin{bmatrix} G^0_{\beta'i'}(\mathbf{x}'',\mathbf{x}) & \varepsilon^0_{\beta'i'j'}(\mathbf{x}'',\mathbf{x}) \\ \varepsilon^0_{\alpha'\beta'i'}(\mathbf{x}'',\mathbf{x}) & E^0_{\alpha'\beta'i'j'}(\mathbf{x}'',\mathbf{x}) \end{bmatrix} \begin{bmatrix} \delta\rho(\mathbf{x})\omega^2\delta_{ij} & 0 \\ 0 & \delta c_{ijkl}(\mathbf{x}) \end{bmatrix} \begin{bmatrix} G_{ja'}(\mathbf{x},\mathbf{x}') \\ \varepsilon_{kla'}(\mathbf{x},\mathbf{x}') \end{bmatrix} dV, \tag{6}$$

where $G^0_{\beta'i'}(\mathbf{x}'',\mathbf{x})$, $\varepsilon^0_{\alpha'\beta'i'}(\mathbf{x}'',\mathbf{x})$, $E^0_{\alpha'\beta'i'j'}(\mathbf{x}'',\mathbf{x})$ are Green's function of the homogeneous reference medium and its derivatives.

For the convenience of subsequent discussion, we rewrite Eq. (6) in a more compact form:

$$\Psi(\mathbf{x}''-\mathbf{x}') = \Psi^0(\mathbf{x}''-\mathbf{x}') + \iiint_{V(\mathbf{x})} \Gamma(\mathbf{x}''-\mathbf{x})\Pi(\mathbf{x})\Psi(\mathbf{x}-\mathbf{x}')d^3\mathbf{x}, \tag{7}$$

where

$$\Psi(\mathbf{x}''-\mathbf{x}') = \begin{bmatrix} G_{11} & G_{12} & G_{13} \\ G_{21} & G_{22} & G_{23} \\ G_{31} & G_{32} & G_{33} \\ G_{11,1} & G_{12,1} & G_{13,1} \\ G_{21,2} & G_{22,2} & G_{23,2} \\ G_{31,3} & G_{32,3} & G_{33,3} \\ G_{21,3}+G_{31,2} & G_{22,3}+G_{32,2} & G_{23,3}+G_{33,2} \\ G_{11,3}+G_{31,1} & G_{12,3}+G_{32,1} & G_{13,3}+G_{33,1} \\ G_{21,1}+G_{11,2} & G_{22,1}+G_{12,2} & G_{23,1}+G_{13,2} \end{bmatrix}, \quad \Psi^0(\mathbf{x}''-\mathbf{x}') = \begin{bmatrix} G^0_{11} & G^0_{12} & G^0_{13} \\ G^0_{21} & G^0_{22} & G^0_{23} \\ G^0_{31} & G^0_{32} & G^0_{33} \\ G^0_{11,1} & G^0_{12,1} & G^0_{13,1} \\ G^0_{21,2} & G^0_{22,2} & G^0_{23,2} \\ G^0_{31,3} & G^0_{32,3} & G^0_{33,3} \\ G^0_{21,3}+G^0_{31,2} & G^0_{22,3}+G^0_{32,2} & G^0_{23,3}+G^0_{33,2} \\ G^0_{11,3}+G^0_{31,1} & G^0_{12,3}+G^0_{32,1} & G^0_{13,3}+G^0_{33,1} \\ G^0_{21,1}+G^0_{11,2} & G^0_{22,1}+G^0_{12,2} & G^0_{23,1}+G^0_{13,2} \end{bmatrix}, \tag{8}$$

$$\Pi(\mathbf{x}) = \begin{bmatrix} \delta\rho\omega^2 & 0 & 0 & 0 & 0 & 0 & 0 & 0 & 0 \\ 0 & \delta\rho\omega^2 & 0 & 0 & 0 & 0 & 0 & 0 & 0 \\ 0 & 0 & \delta\rho\omega^2 & 0 & 0 & 0 & 0 & 0 & 0 \\ 0 & 0 & 0 & \delta c_{11} & \delta c_{12} & \delta c_{13} & \delta c_{14} & \delta c_{15} & \delta c_{16} \\ 0 & 0 & 0 & \delta c_{12} & \delta c_{22} & \delta c_{23} & \delta c_{24} & \delta c_{25} & \delta c_{26} \\ 0 & 0 & 0 & \delta c_{13} & \delta c_{23} & \delta c_{33} & \delta c_{34} & \delta c_{35} & \delta c_{36} \\ 0 & 0 & 0 & \delta c_{14} & \delta c_{24} & \delta c_{34} & \delta c_{44} & \delta c_{45} & \delta c_{46} \\ 0 & 0 & 0 & \delta c_{15} & \delta c_{25} & \delta c_{35} & \delta c_{45} & \delta c_{55} & \delta c_{56} \\ 0 & 0 & 0 & \delta c_{16} & \delta c_{26} & \delta c_{36} & \delta c_{46} & \delta c_{56} & \delta c_{66} \end{bmatrix}, \tag{9}$$

$$\Gamma(\mathbf{x}''-\mathbf{x}) = \begin{bmatrix} \mathbf{A} & \mathbf{B} \\ \mathbf{B}^T & \mathbf{D} \end{bmatrix}, \tag{10}$$

and

$$\mathbf{A} = \begin{bmatrix} G^0_{11} & G^0_{12} & G^0_{13} \\ G^0_{21} & G^0_{22} & G^0_{23} \\ G^0_{31} & G^0_{32} & G^0_{33} \end{bmatrix}, \quad \mathbf{B} = \begin{bmatrix} G^0_{11,1} & G^0_{12,2} & G^0_{13,3} & G^0_{12,3}+G^0_{13,2} & G^0_{11,3}+G^0_{13,1} & G^0_{12,1}+G^0_{11,2} \\ G^0_{21,1} & G^0_{22,2} & G^0_{23,3} & G^0_{22,3}+G^0_{23,2} & G^0_{21,3}+G^0_{23,1} & G^0_{22,1}+G^0_{21,2} \\ G^0_{31,1} & G^0_{32,2} & G^0_{33,3} & G^0_{32,3}+G^0_{33,2} & G^0_{31,3}+G^0_{33,1} & G^0_{32,1}+G^0_{31,2} \end{bmatrix}, \tag{11}$$

$$\mathbf{D} = \begin{bmatrix} G^0_{11,11} & G^0_{12,21} & G^0_{13,31} & G^0_{12,31}+G^0_{13,21} & G^0_{11,31}+G^0_{13,11} & G^0_{12,11}+G^0_{11,21} \\ G^0_{21,12} & G^0_{22,22} & G^0_{23,32} & G^0_{22,32}+G^0_{23,22} & G^0_{21,32}+G^0_{23,12} & G^0_{22,12}+G^0_{21,22} \\ G^0_{31,13} & G^0_{32,23} & G^0_{33,33} & G^0_{32,33}+G^0_{33,23} & G^0_{31,33}+G^0_{33,13} & G^0_{32,13}+G^0_{31,23} \\ G^0_{21,13}+G^0_{31,12} & G^0_{22,23}+G^0_{32,22} & G^0_{23,33}+G^0_{33,32} & G^0_{22,33}+G^0_{33,22}+2G^0_{23,23} & G^0_{21,33}+G^0_{23,13}+G^0_{31,32}+G^0_{33,12} & G^0_{21,23}+G^0_{22,13}+G^0_{31,22}+G^0_{32,12} \\ G^0_{11,31}+G^0_{31,11} & G^0_{12,32}+G^0_{32,12} & G^0_{13,33}+G^0_{33,13} & G^0_{12,33}+G^0_{32,31}+G^0_{13,23}+G^0_{33,21} & G^0_{11,33}+G^0_{33,11}+2G^0_{13,13} & G^0_{12,13}+G^0_{11,23}+G^0_{32,11}+G^0_{31,21} \\ G^0_{11,12}+G^0_{21,11} & G^0_{22,21}+G^0_{12,22} & G^0_{23,31}+G^0_{13,32} & G^0_{12,32}+G^0_{22,31}+G^0_{13,22}+G^0_{23,21} & G^0_{21,31}+G^0_{11,32}+G^0_{23,11}+G^0_{13,12} & G^0_{11,22}+G^0_{22,11}+2G^0_{12,12} \end{bmatrix}. \quad (12)$$

It is noted the propagator $\Gamma$ is a symmetric matrix, i.e., $\Gamma^T = \Gamma$. The fluctuation of the mass density $\delta\rho$ and that of the elastic stiffness $\delta c_{ij}$ appeared in $\Pi(\mathbf{x})$ describe the deviation of the properties of the specific crystallite from those of a homogeneous reference medium, both of which are treated as random variables here. In particular, the elastic stiffness fluctuation $\delta c_{ij}$ is also a function of the crystallographic orientation of the specific crystallite, see Appendix A.

As rigorously analyzed in [51], the integral in the second term on the righthand side of Eq. (7) has $\delta$ singularities due to the second order derivative of the homogeneous Green's function. By introducing the concept of shape-dependent principal value [51], the correct definition of the matrix $\mathbf{D}$ is given by:

$$\mathbf{D} := P.S.\mathbf{D} - \tilde{\mathbf{S}}\delta(\mathbf{x}'' - \mathbf{x}), \quad (13)$$

where $P.S.\ \mathbf{D}$ denotes the shape-dependent principal value of the Green tensor, $\tilde{\mathbf{s}}$ is the singularity of the Green tensor, which is also dependent on the statistic characteristics of the inhomogeneities. For polycrystals, the expression of the singularity tensor $\tilde{\mathbf{s}}$ strongly depends on the geometry of microstructures, such as the grain shape (equiaxed, elongated, or triaxial ellipsoidal) and textures. Generally speaking, there is no closed form expression for the singularity. However, for the simplest case, i.e., equiaxed grains without preferred crystallographic orientation, the material is macroscopically isotropic, and the singularity tensor has the following explicit expression,

$$\tilde{\mathbf{S}} = \begin{bmatrix} S_{1111} & S_{1221} & S_{1221} & 0 & 0 & 0 \\ S_{1221} & S_{1111} & S_{1221} & 0 & 0 & 0 \\ S_{1221} & S_{1221} & S_{1111} & 0 & 0 & 0 \\ 0 & 0 & 0 & 4S_{2233} & 0 & 0 \\ 0 & 0 & 0 & 0 & 4S_{2233} & 0 \\ 0 & 0 & 0 & 0 & 0 & 4S_{2233} \end{bmatrix}, \quad (14)$$

where

$$S_{1111} = \frac{2\lambda + 7\mu}{15\mu(\lambda + 2\mu)}, \quad S_{1221} = -\frac{\lambda + \mu}{15\mu(\lambda + 2\mu)}, \quad S_{2233} = \frac{3\lambda + 8\mu}{30\mu(\lambda + 2\mu)}. \quad (15)$$

where $\lambda$ and $\mu$ are Lame's constants of the homogeneous reference medium.

From Eq. (15) we can see that $S_{1111} = S_{1221} + 2S_{2233}$, thus $S_{ijkl}$ is an isotropic tensor. This conclusion is consistent with the assumption that the polycrystalline medium is macroscopically isotropic. In analogous to the isotropic elastic stiffness tensor, we can rewrite $S_{ijkl}$ as

$$S_{\alpha ij\beta} = S_1 \delta_{\alpha i}\delta_{j\beta} + S_2(\delta_{\alpha j}\delta_{i\beta} + \delta_{\alpha\beta}\delta_{ij}), \quad (16)$$

where

$$S_1 = -\frac{\lambda + \mu}{15(\lambda + 2\mu)\mu}, \quad S_2 = \frac{3\lambda + 8\mu}{30(\lambda + 2\mu)\mu}. \quad (17)$$

Although the dimension of the singularity tensor is the same as that of the elastic compliance tensor, they are completely different quantities. One can easily verify that the expressions of these two quantities are different.

Correspondingly, $P.S.\ \mathbf{D}$ has the following form:

$$P.S.\mathbf{D} = \begin{bmatrix} P.S.E^0_{1111} & P.S.E^0_{1221} & P.S.E^0_{1331} & 2E^0_{1231} & 2E^0_{1131} & 2E^0_{1211} \\ P.S.E^0_{2112} & P.S.E^0_{2222} & P.S.E^0_{2332} & 2E^0_{2232} & 2E^0_{2132} & 2E^0_{2212} \\ P.S.E^0_{3113} & P.S.E^0_{3223} & P.S.E^0_{3333} & 2E^0_{3233} & 2E^0_{3133} & 2E^0_{3213} \\ 2E^0_{2113} & 2E^0_{2223} & 2E^0_{2333} & P.S.4E^0_{2233} & 4E^0_{2133} & 4E^0_{2123} \\ 2E^0_{1131} & 2E^0_{1232} & 2E^0_{1333} & 4E^0_{1233} & P.S.4E^0_{1133} & 4E^0_{1213} \\ 2E^0_{1112} & 2E^0_{2221} & 2E^0_{2331} & 4E^0_{1232} & 4E^0_{2131} & P.S.4E^0_{1122} \end{bmatrix}, \quad (18)$$

Finally, we obtain the correct definition of the matrix $\mathbf{\Gamma}(\mathbf{x}'' - \mathbf{x})$ by introducing its shape-dependent principal value $P.S.\mathbf{\Gamma}(\mathbf{x}'' - \mathbf{x})$, i.e.,

$$\mathbf{\Gamma}(\mathbf{x}'' - \mathbf{x}) := P.S.\mathbf{\Gamma}(\mathbf{x}'' - \mathbf{x}) - \mathbf{S}\delta(\mathbf{x}'' - \mathbf{x}), \quad (19)$$

where

$$P.S.\mathbf{\Gamma}(\mathbf{x}'' - \mathbf{x}) = \begin{bmatrix} \mathbf{A} & \mathbf{B} \\ \mathbf{B}^T & P.S.\mathbf{D} \end{bmatrix}, \quad \mathbf{S} = \begin{bmatrix} \mathbf{0} & \mathbf{0} \\ \mathbf{0} & \tilde{\mathbf{S}} \end{bmatrix}. \quad (20)$$

Substitution of (19) into (7) yields:

$$\mathbf{\Psi}(\mathbf{x}'' - \mathbf{x}') = \mathbf{\Psi}^0(\mathbf{x}'' - \mathbf{x}') + \iiint_{V(\mathbf{x})} [P.S.\mathbf{\Gamma}(\mathbf{x}'' - \mathbf{x}) - \mathbf{S}\delta(\mathbf{x}'' - \mathbf{x})] \mathbf{\Pi}(\mathbf{x}) \mathbf{\Psi}(\mathbf{x} - \mathbf{x}') d^3\mathbf{x}. \quad (21)$$

Invoking the definition of the Dirac-δ function, we get

$$\mathbf{\Psi}(\mathbf{x}'' - \mathbf{x}') = \mathbf{\Psi}^0(\mathbf{x}'' - \mathbf{x}') + \iiint_{V(\mathbf{x})} P.S.\mathbf{\Gamma}(\mathbf{x}'' - \mathbf{x}) \mathbf{\Pi}(\mathbf{x}) \mathbf{\Psi}(\mathbf{x} - \mathbf{x}') d^3\mathbf{x} - \mathbf{S}\mathbf{\Pi}(\mathbf{x}'')\mathbf{\Psi}(\mathbf{x}'' - \mathbf{x}'). \quad (22)$$

Introducing the renormalized field variable defined by

$$\mathbf{\Phi}(\mathbf{x}'' - \mathbf{x}') = \mathbf{\Psi}(\mathbf{x}'' - \mathbf{x}') + \mathbf{S}\mathbf{\Pi}(\mathbf{x}'')\mathbf{\Psi}(\mathbf{x}'' - \mathbf{x}'), \quad (23)$$

the integral equation (22) is rewritten as:

$$\mathbf{\Phi}(\mathbf{x}'' - \mathbf{x}') = \mathbf{\Psi}^0(\mathbf{x}'' - \mathbf{x}') + \iiint_{V(\mathbf{x})} P.S.\mathbf{\Gamma}(\mathbf{x}'' - \mathbf{x}) \mathbf{\Xi}(\mathbf{x}) \mathbf{\Phi}(\mathbf{x} - \mathbf{x}') d^3\mathbf{x}, \quad (24)$$

where

$$\mathbf{\Xi}(\mathbf{x}) = \mathbf{\Pi}(\mathbf{x})[\mathbf{I} + \mathbf{S}\mathbf{\Pi}(\mathbf{x})]^{-1}. \quad (25)$$

Equation (24) is the renormalized integral equation for the renormalized field $\mathbf{\Phi}(\mathbf{x}''-\mathbf{x}')$. In this work, we consider polycrystalline materials with general anisotropic crystallites, correspondingly, the matrix $\mathbf{\Xi}(\mathbf{x})$ is given by:

$$\mathbf{\Xi}(\mathbf{x}) = \begin{bmatrix} \delta\rho\omega^2 & 0 & 0 & 0 & 0 & 0 & 0 & 0 & 0 \\ 0 & \delta\rho\omega^2 & 0 & 0 & 0 & 0 & 0 & 0 & 0 \\ 0 & 0 & \delta\rho\omega^2 & 0 & 0 & 0 & 0 & 0 & 0 \\ 0 & 0 & 0 & \Xi_{11} & \Xi_{12} & \Xi_{13} & \Xi_{14} & \Xi_{15} & \Xi_{16} \\ 0 & 0 & 0 & \Xi_{12} & \Xi_{22} & \Xi_{23} & \Xi_{24} & \Xi_{25} & \Xi_{26} \\ 0 & 0 & 0 & \Xi_{13} & \Xi_{23} & \Xi_{33} & \Xi_{34} & \Xi_{35} & \Xi_{36} \\ 0 & 0 & 0 & \Xi_{14} & \Xi_{24} & \Xi_{34} & \Xi_{44} & \Xi_{45} & \Xi_{46} \\ 0 & 0 & 0 & \Xi_{15} & \Xi_{25} & \Xi_{35} & \Xi_{45} & \Xi_{55} & \Xi_{56} \\ 0 & 0 & 0 & \Xi_{16} & \Xi_{26} & \Xi_{36} & \Xi_{46} & \Xi_{56} & \Xi_{66} \end{bmatrix}. \quad (26)$$

Similar to the elastic stiffness fluctuation $\delta c_{ij}$, the renormalized elastic variable $\Xi_{pq}$ is also dependent on the orientation of each grain, i.e., it is a function of the Euler angles. The governing equation for the resemble average of the stochastic variable, i.e., the coherent response of the scattered field, can be obtained using the Feynman diagram technique, for which the general procedure is presented in the author's previous work [51]. If further invoking the first-order-smoothing approximation [44, 51, 53-54], then we obtain Dyson's equation:

$$\langle \Phi_{af}(\mathbf{x}'' - \mathbf{x}') \rangle = \Psi^0_{af}(\mathbf{x}'' - \mathbf{x}') + \iiint_{V(\mathbf{y})} P.S.\Gamma_{ab}(\mathbf{x}'' - \mathbf{y}) \iiint_{V(\mathbf{x})} P.S.\Gamma_{cd}(\mathbf{y} - \mathbf{x}) \langle \Xi_{bc}(\mathbf{y}) \Xi_{de}(\mathbf{x}) \rangle \langle \Phi_{ef}(\mathbf{x} - \mathbf{x}') \rangle d^3\mathbf{x}, \quad (27)$$

For a statistically homogeneous medium, the two-point correlation function is given by:

$$\langle \Xi_{ab}(\mathbf{y}) \Xi_{cd}(\mathbf{x}) \rangle = \langle \Xi_{ab} \Xi_{cd} \rangle P(\mathbf{y} - \mathbf{x}), \quad (28)$$

where the ensemble-average of the covariance of the renormalized elastic quantity $\langle \Xi_{ab}(\mathbf{y})\Xi_{cd}(\mathbf{x})\rangle$ are separated into orientation dependent part $\langle \Xi_{ab}\Xi_{cd}\rangle$ and the spatial dependent part $P(\mathbf{y}-\mathbf{x})$. This conclusion is hold when the random medium is statistically homogeneous and the ergodic assumption is satisfied. The ensemble average of $\langle \Xi_{ab}\Xi_{cd}\rangle$ is taken on the entire SO(3) group. The invariant integral measure (Haar measure) on SO(3) group can be calculated from the biinvariant metric and the corresponding left-invariant volumetric 3-form. If the SO(3) group is parameterized by the Euler angles, the normalized volumetric element/measure is $(8\pi^2)^{-1}\sin\theta d\varphi \wedge d\theta \wedge d\psi$ [44, 55], thus $\langle \Xi_{ab}\Xi_{cd}\rangle$ has the following expression:

$$\langle \Xi_{ab}(\varphi,\theta,\psi)\Xi_{cd}(\varphi,\theta,\psi)\rangle = \frac{1}{8\pi^2}\int_0^{2\pi}d\varphi\int_0^{\pi}d\theta\int_0^{2\pi}d\psi \Xi_{ab}(\varphi,\theta,\psi)\Xi_{cd}(\varphi,\theta,\psi)\sin\theta, \quad (29)$$

$P(\mathbf{y}-\mathbf{x})$ is called spatial autocorrelation function (SAF). The SAF of polycrystalline materials are extensively studied by Stanke [31]. It is shown the exponential functions can best fit the experimentally measured SAF. Consequently, in this work, we use the exponential correlation function to describe the statistical characteristics of polycrystals. Furthermore, we consider polycrystalline materials with equiaxed grains, thus the bulk material is macroscopically isotropic, and the exponential correlation function is dependent on the magnitude of the distance of two points only:

$$P(\mathbf{y}-\mathbf{x}) = e^{-\frac{|\mathbf{y}-\mathbf{x}|}{a}}, \quad (30)$$

where $a$ is the correlation length. It is generally considered as the average radius of the grains.

The Fourier transform of SAF represents the power spectrum of the medium fluctuation, which is given by:

$$\tilde{P}(\mathbf{k}) = \frac{8\pi a^3}{(1+k^2a^2)^2}, \quad (31)$$

where $k$ is the magnitude of the wavenumber.

As pointed out in [51], we can always choose the properties of the reference medium such that the first order moment vanishes. This choice also ensures the fastest convergence rate of the multiple-scattering series. According to the ergodic assumption, the ensemble average is equivalent to the volumetric average. For the case of untextured polycrystals, the volumetric average is further equal to the average over the whole space of possible crystallographic orientations, thus we get:

$$\langle \Xi_{ab}(\mathbf{x})\rangle = 0, \quad (32)$$

and

$$<\Xi_{ab}> = \frac{1}{8\pi^2}\int_0^{2\pi}d\varphi\int_0^{\pi}d\theta\int_0^{2\pi}d\psi \Xi_{ab}(\varphi,\theta,\psi)\sin\theta. \quad (33)$$

Eq. (33) gives 36 equations for the unknow elastic properties of the reference medium. Due to the isotropy of the reference medium, it has only two independent parameters, i.e., the Lame constants. It is shown numerically that only two of the 36 equations are independent, which exactly form a system of well-defined equations. The Lame constants can be obtained by solving any two independent equations of the 36 equations, here we use:

$$\langle \Xi_{11}\rangle = 0: \quad \frac{1}{8\pi^2}\int_0^{2\pi}d\varphi\int_0^{\pi}d\theta\int_0^{2\pi}d\psi \Xi_{11}(\varphi,\theta,\psi)\sin\theta = 0, \quad (34)$$

$$\langle \Xi_{44}\rangle = 0: \quad \frac{1}{8\pi^2}\int_0^{2\pi}d\varphi\int_0^{\pi}d\theta\int_0^{2\pi}d\psi \Xi_{44}(\varphi,\theta,\psi)\sin\theta = 0. \quad (35)$$

In this work, we only consider single phase polycrystals, so the density is a constant. But we need to mention that the general theoretical framework is suitable for multiphase polycrystals for which both the density and elastic stiffness are random variables. The elastic stiffness constants of the homogeneous reference material obtained by solving Eqs. (34) and (35) are largely different from the Voigt-averaged values. In Sections III and IV we will see that the Voigt average scheme gives the upper bond of the quasi-static limit of the coherent wave velocities. This conclusion is in agreement with that given by other works [31, 50].

Eq. (27) is a system of integral equations of convolution type, it is most conveniently solved by the Fourier transformation technique. In this work, we use the following Fourier transform pair:

$$\tilde{\Psi}(\mathbf{k}) = \iiint_{V(\mathbf{x})} \Psi(\mathbf{x}) e^{-i\mathbf{k}\cdot\mathbf{x}} d^3\mathbf{x}, \quad \Psi(\mathbf{x}) = \frac{1}{8\pi^3} \iiint_{V(\mathbf{k})} \tilde{\Psi}(\mathbf{k}) e^{i\mathbf{k}\cdot\mathbf{x}} d^3\mathbf{k} , \tag{36}$$

Applying Fourier transform to the renormalized Dyson's equation (27), and considering (30) and (31), we get:

$$\langle \tilde{\Phi}_{af}(\mathbf{k}) \rangle = \tilde{\Psi}^0_{af}(\mathbf{k}) + P.S.\tilde{\Gamma}_{ab}(\mathbf{k}) \langle \Xi_{bc}\Xi_{de} \rangle \left[ \frac{1}{8\pi^3} \iiint_{V(\mathbf{s})} P.S.\tilde{\Gamma}_{cd}(\mathbf{s}) \tilde{P}(\mathbf{k}-\mathbf{s}) d^3\mathbf{s} \right] \langle \tilde{\Phi}_{ef}(\mathbf{k}) \rangle, \tag{37}$$

Multiplying both sides of Eq. (37) by $\left[ P.S.\tilde{\Gamma}_{ga}(\mathbf{k}) \right]^{-1}$, we get:

$$\left[ P.S.\tilde{\Gamma}_{ga}(\mathbf{k}) \right]^{-1} \langle \tilde{\Phi}_{af}(\mathbf{k}) \rangle = \left[ P.S.\tilde{\Gamma}_{ga}(\mathbf{k}) \right]^{-1} \tilde{\Psi}^0_{af}(\mathbf{k}) + \langle \Xi_{gc}\Xi_{de} \rangle \left[ \frac{1}{8\pi^3} \iiint_{V(\mathbf{s})} P.S.\tilde{\Gamma}_{cd}(\mathbf{s}) \tilde{P}(\mathbf{k}-\mathbf{s}) d^3\mathbf{s} \right] \langle \tilde{\Phi}_{ef}(\mathbf{k}) \rangle, \tag{38}$$

Rearranging this equation, we have:

$$\left\{ \left[ P.S.\tilde{\Gamma}_{ge}(\mathbf{k}) \right]^{-1} - \langle \Xi_{gc}\Xi_{de} \rangle \left[ \frac{1}{8\pi^3} \iiint_{V(\mathbf{s})} P.S.\tilde{\Gamma}_{cd}(\mathbf{s}) \tilde{P}(\mathbf{k}-\mathbf{s}) d^3\mathbf{s} \right] \right\} \langle \tilde{\Phi}_{ef}(\mathbf{k}) \rangle = \left[ P.S.\tilde{\Gamma}_{ga}(\mathbf{k}) \right]^{-1} \tilde{\Psi}^0_{af}(\mathbf{k}), \tag{39}$$

The ensemble averaged response in the frequency-wavenumber domain can be solved by:

$$\langle \tilde{\Phi}_{hf}(\mathbf{k}) \rangle = \left\{ \left[ P.S.\tilde{\Gamma}_{hg}(\mathbf{k}) \right]^{-1} - \langle \Xi_{hc}\Xi_{dg} \rangle \left[ \frac{1}{8\pi^3} \iiint_{V(\mathbf{s})} P.S.\tilde{\Gamma}_{cd}(\mathbf{s}) \tilde{P}(\mathbf{k}-\mathbf{s}) d^3\mathbf{s} \right] \right\}^{-1} \left[ P.S.\tilde{\Gamma}_{ga}(\mathbf{k}) \right]^{-1} \tilde{\Psi}^0_{af}(\mathbf{k}), \tag{40}$$

Simultaneously we obtain the dispersion equation:

$$\det \left\{ \left[ P.S.\tilde{\Gamma}_{ge}(\mathbf{k}) \right]^{-1} - \langle \Xi_{gc}\Xi_{de} \rangle \left[ \frac{1}{8\pi^3} \iiint_{V(\mathbf{s})} P.S.\tilde{\Gamma}_{cd}(\mathbf{s}) \tilde{P}(\mathbf{k}-\mathbf{s}) d^3\mathbf{s} \right] \right\} = 0 , \tag{41}$$

The solution in the frequency domain is given by:

$$\langle \Phi_{hf}(\mathbf{x},\mathbf{x}',\omega) \rangle = \frac{1}{8\pi^3} \iiint_{V(\mathbf{k})} \left\{ \left[ P.S.\tilde{\Gamma}_{hg}(\mathbf{k}) \right]^{-1} - \langle \Xi_{hc}\Xi_{dg} \rangle \left[ \frac{1}{8\pi^3} \iiint_{V(\mathbf{s})} P.S.\tilde{\Gamma}_{cd}(\mathbf{s}) \tilde{P}(\mathbf{k}-\mathbf{s}) d^3\mathbf{s} \right] \right\}^{-1} \left[ P.S.\tilde{\Gamma}_{ga}(\mathbf{k}) \right]^{-1} \tilde{\Psi}^0_{af}(\mathbf{k}) e^{i\mathbf{k}\cdot(\mathbf{x}-\mathbf{x}')} d^3\mathbf{k} , \tag{42}$$

If the source is a time varying signal $\mathbf{F}(t)$, its spectrum can be obtained by the following time domain Fourier transform pair:

$$\tilde{F}(\omega) = \int_{-\infty}^{+\infty} F(t) e^{i\omega t} dt , \quad F(t) = \frac{1}{2\pi} \int_{-\infty}^{+\infty} \tilde{F}(\omega) e^{-i\omega t} d\omega , \tag{43}$$

and the complete wavefield in spacetime is given by:

$$\langle \Phi_{hf}(\mathbf{x},\mathbf{x}',t) \rangle =$$
$$\frac{1}{2\pi} \int_{-\infty}^{+\infty} \tilde{F}(\omega) \left( \frac{1}{8\pi^3} \iiint_{V(\mathbf{k})} \left\{ \left[ P.S.\tilde{\Gamma}_{hg}(\mathbf{k}) \right]^{-1} - \langle \Xi_{hc}\Xi_{dg} \rangle \left[ \frac{1}{8\pi^3} \iiint_{V(\mathbf{s})} P.S.\tilde{\Gamma}_{cd}(\mathbf{s}) \tilde{P}(\mathbf{k}-\mathbf{s}) d^3\mathbf{s} \right] \right\}^{-1} \left[ P.S.\tilde{\Gamma}_{ga}(\mathbf{k}) \right]^{-1} \tilde{\Psi}^0_{af}(\mathbf{k}) e^{i\mathbf{k}\cdot(\mathbf{x}-\mathbf{x}')} d^3\mathbf{k} \right) e^{-i\omega t} d\omega, \tag{44}$$

Eq. (44) gives an explicit expression for the mean wavefield induced by a general point source. It has significant implications for applications in ultrasonic nondestructive evaluation technologies. For instance, it can be used to develop grain noise models when integrated with transducer transfer functions. In this work, we only focus on the dispersion and attenuation behavior of a plane wave component. In a statistically isotropic medium, the dispersion behavior of a plane wave is independent of its propagation direction. Without loss of generality, we consider a plane wave propagating along the $x_3$ axis, the wavevector $\mathbf{k} = [0,0,k]$, the coefficient matrix and the dispersion equations of longitudinal and transverse waves can be expressed explicitly as follows:

$$\left[P.S.\tilde{\Gamma}_{ge}(\mathbf{k})\right]^{-1} - \langle \Xi_{gc}\Xi_{de}\rangle\left[\frac{1}{8\pi^3}\iiint_{V(\mathbf{s})} P.S.\tilde{\Gamma}_{cd}(\mathbf{s})\tilde{P}(\mathbf{k}-\mathbf{s})d^3\mathbf{s}\right] = \begin{bmatrix} M_{11} & 0 & 0 & 0 & 0 & 0 & 0 & M_{18} & 0 \\ 0 & M_{22} & 0 & 0 & 0 & 0 & M_{27} & 0 & 0 \\ 0 & 0 & M_{33} & M_{34} & M_{35} & M_{36} & 0 & 0 & 0 \\ 0 & 0 & M_{34} & M_{44} & M_{45} & M_{46} & 0 & 0 & 0 \\ 0 & 0 & M_{35} & M_{45} & M_{55} & M_{56} & 0 & 0 & 0 \\ 0 & 0 & M_{36} & M_{46} & M_{56} & M_{66} & 0 & 0 & 0 \\ 0 & M_{27} & 0 & 0 & 0 & 0 & M_{77} & 0 & 0 \\ M_{18} & 0 & 0 & 0 & 0 & 0 & 0 & M_{88} & 0 \\ 0 & 0 & 0 & 0 & 0 & 0 & 0 & 0 & M_{99} \end{bmatrix}, \quad (45)$$

The non-vanishing elements of the matrix **M** are given by:

$$M_{11} = \mu(k^2 - k_T^2) - K_{44}k^2, \quad M_{33} = (\lambda + 2\mu)(k^2 - k_L^2) - K_{11}k^2, \quad M_{34} = -K_{12}ik, \quad M_{36} = -K_{11}ik, \quad M_{18} = -K_{44}ik, \quad (46a)$$

$$M_{44} = K_{11} - [<\Xi_{11}^2> + <\Xi_{12}^2>]\Sigma_{44} - 2<\Xi_{11}\Xi_{12}>\Sigma_{45} - 2[<\Xi_{11}\Xi_{13}> + <\Xi_{12}\Xi_{13}>]\Sigma_{46} - <\Xi_{13}^2>\Sigma_{66} - [<\Xi_{14}^2> + <\Xi_{15}^2>]\Sigma_{77} - <\Xi_{16}^2>\Sigma_{99}, \quad (46b)$$

$$M_{45} = K_{12} - [<\Xi_{11}\Xi_{12}> + <\Xi_{12}\Xi_{22}>]\Sigma_{44} - [<\Xi_{12}^2> + <\Xi_{11}\Xi_{22}>]\Sigma_{45} - [<\Xi_{12}\Xi_{13}> + <\Xi_{13}\Xi_{22}> + <\Xi_{11}\Xi_{23}> + <\Xi_{12}\Xi_{23}>]\Sigma_{46}$$
$$- <\Xi_{13}\Xi_{23}>\Sigma_{66} - [<\Xi_{14}\Xi_{24}> + <\Xi_{15}\Xi_{25}>]\Sigma_{77} - <\Xi_{16}\Xi_{26}>\Sigma_{99}, \quad (46c)$$

$$M_{46} = K_{12} - [<\Xi_{11}\Xi_{13}> + <\Xi_{12}\Xi_{23}>]\Sigma_{44} - [<\Xi_{12}\Xi_{13}> + <\Xi_{11}\Xi_{23}>]\Sigma_{45} - [<\Xi_{13}^2> + <\Xi_{13}\Xi_{23}> + <\Xi_{11}\Xi_{33}> + <\Xi_{12}\Xi_{33}>]\Sigma_{46}$$
$$- <\Xi_{13}\Xi_{33}>\Sigma_{66} - [<\Xi_{14}\Xi_{34}> + <\Xi_{15}\Xi_{35}>]\Sigma_{77} - <\Xi_{16}\Xi_{36}>\Sigma_{99}, \quad (46d)$$

$$M_{55} = K_{11} - [<\Xi_{12}^2> + <\Xi_{22}^2>]\Sigma_{44} - 2<\Xi_{12}\Xi_{22}>\Sigma_{45} - 2[<\Xi_{12}\Xi_{23}> + <\Xi_{22}\Xi_{23}>]\Sigma_{46} - <\Xi_{23}^2>\Sigma_{66} - [<\Xi_{24}^2> + <\Xi_{25}^2>]\Sigma_{77} - <\Xi_{26}^2>\Sigma_{99}, \quad (46e)$$

$$M_{56} = K_{12} - [<\Xi_{12}\Xi_{13}> + <\Xi_{22}\Xi_{23}>]\Sigma_{44} - [<\Xi_{13}\Xi_{22}> + <\Xi_{12}\Xi_{23}>]\Sigma_{45} - [<\Xi_{23}^2> + <\Xi_{13}\Xi_{23}> + <\Xi_{12}\Xi_{33}> + <\Xi_{22}\Xi_{33}>]\Sigma_{46}$$
$$- <\Xi_{23}\Xi_{33}>\Sigma_{66} - [<\Xi_{24}\Xi_{34}> + <\Xi_{25}\Xi_{35}>]\Sigma_{77} - <\Xi_{26}\Xi_{36}>\Sigma_{99}, \quad (46f)$$

$$M_{66} = K_{11} - [<\Xi_{13}^2> + <\Xi_{23}^2>]\Sigma_{44} - 2<\Xi_{13}\Xi_{23}>\Sigma_{45} - 2[<\Xi_{13}\Xi_{33}> + <\Xi_{23}\Xi_{33}>]\Sigma_{46} - <\Xi_{33}^2>\Sigma_{66} - [<\Xi_{34}^2> + <\Xi_{35}^2>]\Sigma_{77} - <\Xi_{36}^2>\Sigma_{99}, \quad (46g)$$

$$M_{77} = K_{44} - [<\Xi_{14}^2> + <\Xi_{24}^2>]\Sigma_{44} - 2<\Xi_{14}\Xi_{24}>\Sigma_{45} - 2[<\Xi_{14}\Xi_{34}> + <\Xi_{24}\Xi_{34}>]\Sigma_{46} - <\Xi_{34}^2>\Sigma_{66} - [<\Xi_{44}^2> + <\Xi_{45}^2>]\Sigma_{77} - <\Xi_{46}^2>\Sigma_{99}, \quad (46h)$$

$$M_{88} = K_{44} - [<\Xi_{15}^2> + <\Xi_{25}^2>]\Sigma_{44} - 2<\Xi_{15}\Xi_{25}>\Sigma_{45} - 2[<\Xi_{15}\Xi_{35}> + <\Xi_{25}\Xi_{35}>]\Sigma_{46} - <\Xi_{35}^2>\Sigma_{66} - [<\Xi_{45}^2> + <\Xi_{55}^2>]\Sigma_{77} - <\Xi_{56}^2>\Sigma_{99}, \quad (46i)$$

$$M_{99} = K_{44} - [<\Xi_{16}^2> + <\Xi_{26}^2>]\Sigma_{44} - 2<\Xi_{16}\Xi_{26}>\Sigma_{45} - 2[<\Xi_{16}\Xi_{36}> + <\Xi_{26}\Xi_{36}>]\Sigma_{46} - <\Xi_{36}^2>\Sigma_{66} - [<\Xi_{46}^2> + <\Xi_{56}^2>]\Sigma_{77} - <\Xi_{66}^2>\Sigma_{99}, \quad (46i)$$

$$M_{22} = M_{11}, \quad M_{35} = M_{34}, \quad M_{27} = M_{18}, \quad (46j)$$

where

$$K_{11} = \frac{3(\lambda + 6\mu)(\lambda + 2\mu)}{3\lambda + 8\mu}, \quad K_{12} = \frac{3(\lambda + \mu)(\lambda + 2\mu)}{3\lambda + 8\mu}, \quad K_{44} = \frac{15(\lambda + 2\mu)\mu}{2(3\lambda + 8\mu)}, \quad K_{11} = K_{12} + 2K_{44}. \quad (47)$$

$$\Sigma_{44}(\mathbf{k}) = S_{1111} - \frac{1}{8\pi^3}\iiint_{V(\mathbf{s})} s_1^2 \tilde{G}_{11}(\mathbf{s})\tilde{P}(\mathbf{k}-\mathbf{s})d^3\mathbf{s}, \quad \Sigma_{45}(\mathbf{k}) = S_{1221} - \frac{1}{8\pi^3}\iiint_{V(\mathbf{s})} s_1 s_2 \tilde{G}_{12}(\mathbf{s})\tilde{P}(\mathbf{k}-\mathbf{s})d^3\mathbf{s}, \quad (48a)$$

$$\Sigma_{46}(\mathbf{k}) = S_{1221} - \frac{1}{8\pi^3}\iiint_{V(\mathbf{s})} s_1 s_3 \tilde{G}_{13}(\mathbf{s})\tilde{P}(\mathbf{k}-\mathbf{s})d^3\mathbf{s}, \quad \Sigma_{66}(\mathbf{k}) = S_{1111} - \frac{1}{8\pi^3}\iiint_{V(\mathbf{s})} s_3^2 \tilde{G}_{33}(\mathbf{s})\tilde{P}(\mathbf{k}-\mathbf{s})d^3\mathbf{s}, \quad (48b)$$

$$\Sigma_{77}(\mathbf{k}) = 4S_{2233} - \frac{1}{8\pi^3}\iiint_{V(\mathbf{s})} [s_2^2 \tilde{G}_{33}(\mathbf{s}) + s_3^2 \tilde{G}_{22}(\mathbf{s}) + 2s_2 s_3 \tilde{G}_{23}(\mathbf{s})]\tilde{P}(\mathbf{k}-\mathbf{s})d^3\mathbf{s}, \quad (48c)$$

$$\Sigma_{99}(\mathbf{k}) = 4S_{2233} - \frac{1}{8\pi^3}\iiint_{V(\mathbf{s})} [s_1^2 \tilde{G}_{22}(\mathbf{s}) + s_2^2 \tilde{G}_{11}(\mathbf{s}) + 2s_1 s_2 \tilde{G}_{12}(\mathbf{s})]\tilde{P}(\mathbf{k}-\mathbf{s})d^3\mathbf{s}, \quad (48d)$$

$$\Sigma_{22} = \Sigma_{11}, \quad \Sigma_{27} = \Sigma_{18}, \quad \Sigma_{35} = \Sigma_{34}, \quad \Sigma_{55} = \Sigma_{44}, \quad \Sigma_{56} = \Sigma_{46}, \quad \Sigma_{88} = \Sigma_{77}, \quad (48e)$$

The dispersion equation for longitudinal coherent waves is:

$$M_{33}(M_{44}M_{56}^2 + M_{55}M_{46}^2 + M_{66}M_{45}^2 - 2M_{45}M_{46}M_{56} - M_{44}M_{55}M_{66})$$
$$+ M_{34}^2(M_{55}M_{66} - M_{56}^2) + M_{35}^2(M_{44}M_{66} - M_{46}^2) + M_{36}^2(M_{44}M_{55} - M_{45}^2) \quad (49)$$
$$+ 2M_{34}M_{35}(M_{46}M_{56} - M_{45}M_{66}) + 2M_{34}M_{36}(M_{45}M_{56} - M_{46}M_{55}) + 2M_{35}M_{36}(M_{45}M_{46} - M_{44}M_{56}) = 0,$$

The dispersion equation for coherent transverse waves is:

$$M_{11}M_{88} - M_{18}^2 = 0, \tag{50}$$

The propagation characteristics of the coherent wave, i.e., the ensemble average of the perturbed field can be described by the complex propagation constant, $k = \text{Re}(k) + i\,\text{Im}(k)$, $\text{Re}(k) = \omega/V$, $\text{Im}(k) = \alpha$, where $\omega$ is the circular frequency, $V$ is the coherent wave velocity, and $\alpha$ is the attenuation coefficient of the coherent wave. The procedure for the calculation of the involved integrals are detailed in [51]. All the integrands in the above integrals decays proportionally to $1/s^2$ when $s \to +\infty$. Consequently, all the infinite integrals are convergent. The complex wavenumber is obtained by searching for the roots of the dispersion equations in the complex $k$-plane. The numerical algorithm is implemented on the platform Compaq Visual Fortran 6.6 for which the powerful IMSL numerical library is integrated.

We introduce the dimensionless quantities in the following discussion. The dimensionless velocity variation, attenuations and frequencies of longitudinal and transverse coherent waves are defined by:

$$\delta \bar{V}_L = \frac{V - V_{0L}}{V_{0L}}, \quad \bar{\alpha}_L = \alpha_L d, \quad \bar{K}_{0L} = k_{0L} d, \tag{51}$$

$$\delta \bar{V}_T = \frac{V - V_{0T}}{V_{0T}}, \quad \bar{\alpha}_T = \alpha_T d, \quad \bar{K}_{0T} = k_{0T} d, \tag{52}$$

where $d$ is the average diameter (or characteristic dimension) of the heterogeneities, $d=2a$, $V_{0L}$ and $V_{0T}$ are the velocities of the longitudinal and transverse waves of the reference material, $k_{0L}$ and $k_{0T}$ are the converted wavenumbers calculated using the longitudinal and transverse waves of the reference medium, i.e., $k_{0L}=\omega/V_{0L}$, $k_{0T}=\omega/V_{0T}$. Since $d$, $V_{0L}$ and $V_{0T}$ are constants, $\bar{K}_{0L}$ and $\bar{K}_{0T}$ can be viewed as dimensionless frequencies. $\alpha_L$ and $\alpha_T$ are the attenuation coefficients of longitudinal and transverse waves.

## III. VALIDATION OF THE NEW MODEL

To validate the accuracy of the new model, we first calculate the velocity and attenuation of several practically important polycrystalline alloys, such as aluminum alloy, iron alloy, 304 stainless, OFHC copper, and Inconel 600. These materials have been widely used in aeronautic industry, oil and gas transmission, and pressured vessels and pipes in nuclear reactors [56-57]. Consequently, nondestructive characterization of these materials has drawn much attentions. Stanke [31] calculated the dispersion and attenuation of Al and Iron using the unified scattering theory. Experimental data for the attenuation of longitudinal waves in 304 stainless steel and OFHC copper are also measured. Pamel [56-57] conducted comprehensive numerical simulations on the propagation and scattering of ultrasonic waves in polycrystalline materials. The experimental data and numerical simulations provides good references to evaluate the accuracy of the new model.

Table 1 shows the elastic stiffness and the density of the materials used for comparison. The materials considered by Stanke [31] and Pamel [56-57] all belong to the body-centered cubic (BCC) symmetry class. The degree of anisotropy of cubic single crystals is described by the following relative anisotropy factor:

$$\varepsilon = \frac{|c_{11} - c_{12} - 2c_{44}|}{c_{11}^0}, \text{ where } c_{11}^0 = \lambda + 2\mu. \tag{53}$$

where $\lambda$ and $\mu$ are the Lame constants of the reference medium, as shown in Tab. 2.

From the material parameters we can see the crystallites in aluminum alloy have relatively low anisotropy, $\varepsilon=0.1$. Crystallites of Iron, OFHC copper and Inconel 600 have intermediate degree of anisotropy, $\varepsilon \approx 0.5$. 304 stainless steel has the strongest grain anisotropy, $\varepsilon \approx 0.7$. The various degrees of grain anisotropy indicate these materials are good examples to validate the new model.

Table 1 Material properties of common metal polycrystals

| Material | Elastic constants (GPa) | Density (kg/m$^3$) | Relative anisotropy factor |
|---|---|---|---|

| | | $C_{11}$ | $C_{12}$ | $C_{44}$ | $\rho$ | $\varepsilon$ |
|---|---|---|---|---|---|---|
| Aluminum | BCC | 103.4 | 57.1 | 28.6 | 2700 | 0.10 |
| Iron | BCC | 219.2 | 136.8 | 109.2 | 7860 | 0.51 |
| 304 SS | BCC | 200.5 | 133.0 | 125.0 | 8010 | 0.70 |
| OFHC Copper | BCC | 168.3 | 121.1 | 75.7 | 8392 | 0.51 |
| Inconel 600 | BCC | 234.6 | 145.4 | 126.2 | 8260 | 0.56 |

The elastic stiffness of the homogeneous reference media are calculated by solving the system of nonlinear equations (34)-(35). The Voigt average elastic moduli are given by:

$$\bar{\lambda} = \frac{1}{15}[c_{11} + c_{22} + c_{33} + 4(c_{12} + c_{13} + c_{23}) - 2(c_{44} + c_{55} + c_{66})],$$
$$\bar{\mu} = \frac{1}{15}[c_{11} + c_{22} + c_{33} - c_{12} - c_{13} - c_{23} + 3(c_{44} + c_{55} + c_{66})],$$
(54)

As pointed out in the introduction, most of the existing scattering theories, including the unified theory developed by Stanke and Kino [30] and Weaver's model [44], use the Voigt average material as the homogeneous reference medium. It is well known the Voigt average properties are calculated based on the equal strain assumption, and thus it always overestimates the elastic stiffness of the materials. Contrary to the Voigt average, the Reuss average is obtained with the equal stress assumption, which gives the lower bond of the elastic stiffness. Shtrikman and Hashin [58] developed a self-consistent method to evaluate the effective elastic properties of polycrystalline materials. In the scattering scenario, the velocities calculated using the predictions of all these homogenization schemes give the quasi-static limits of the coherent wave velocities. To quantitatively compare the difference given by these bonds for various materials, Tab. 2 lists the reference properties calculated from the new theory and the Voigt average properties, and Tab. 3 lists the quasi-static limits of phase velocities predicted by various models. The acronym SFMS stands for the Strong-Fluctuation-Multiple-Scattering theory developed in this work. Using the predictions given by the SFMS theory as a standard, Tab. 2 shows that for Aluminum, the Voigt average approach overestimates the longitudinal velocity by 0.07%, and overestimate 0.23% for transverse velocity. For strongly anisotropic material 304 stainless steel, the relative error for longitudinal and transverse velocity goes up to 2.78% and 7.24%, respectively. The quasi-static limits of longitudinal and transverse waves predicted by the SFMS theory and the unified theory demonstrate excellent agreement for all the materials, the relative differences always lie in 0.08 %, which are negligible. The Reuss bond and the Voigt bond gives reasonable predictions for polycrystals with weakly anisotropic grains, but for materials with strongly anisotropic grains the relative error of the estimates becomes large.

Table 2 Reference velocities of different materials (Unit: m/s)

| Material | Reference property | | Voigt property | | Reference velocity | | Voigt velocity | |
|---|---|---|---|---|---|---|---|---|
| | $\lambda$(GPa) | $\mu$ (GPa) | $\bar{\lambda}$ (GPa) | $\bar{\mu}$ (GPa) | $V_{0L}$ | $V_{0T}$ | $\bar{V}_L$ | $\bar{V}_T$ |
| Aluminum | 55.01 | 26.30 | 54.92 | 26.42 | 6313.13 | 3121.02 | 6317.52 | 3128.13 |
| Iron | 114.41 | 75.47 | 109.60 | 82.00 | 5810.30 | 3098.67 | 5899.93 | 3229.95 |
| 304 SS | 104.99 | 76.95 | 96.50 | 88.50 | 5685.14 | 3099.47 | 5843.36 | 3323.96 |
| OFHC copper | 104.49 | 49.12 | 100.26 | 54.86 | 4915.03 | 2419.34 | 5002.14 | 2556.79 |
| Inconel 600 | 118.93 | 85.19 | 112.76 | 93.56 | 5918.23 | 3211.47 | 6025.37 | 3365.54 |

Table 3. Quasi-static limits of phase velocities predicted by different models (Unit: m/s)

| | Reuss | Shtrikman | Unified theory | SFMS | Hashin | Voight |
|---|---|---|---|---|---|---|
| Al, $V_L$ | 6308.3 | 6314.0 | 6314.5 | 6313.13 | 6314.5 | 6317.52 |
| Al, $V_T$ | 3113.5 | 3122.3 | 3122.9 | 3121.02 | 3123.0 | 3128.13 |
| Fe, $V_L$ | 5661.9 | 5764.0 | 5805.7 | 5810.30 | 5810.8 | 5899.93 |
| Fe, $V_T$ | 2892.8 | 3040.3 | 3099.3 | 3098.67 | 3106.5 | 3229.95 |

The velocity and attenuation of longitudinal and transverse waves in Aluminum, Iron and Inconel 600 predicted by the SFMS theory in the whole frequency range are shown in Fig. 2. From the numerical results we can observe the dispersion and attenuation behaviors of these materials exhibit several common features: 1) The dispersion curves of both L and T waves start from the velocities of the homogeneous reference materials. This shows that the quasi-static limits of the coherent waves naturally converge to the velocities of the homogeneous reference materials; 2) At low frequencies $k_0 d < 1$, the dispersion is nearly negligible, and the dimensionless attenuation increases with dimensionless frequency following a power law; 3) At intermediate frequencies, the dispersion slightly increases with frequency, meanwhile, a second propagation mode starts to appear. The magnitude of the dispersion strongly depends on the degree of grain anisotropy, the stronger the anisotropy, the larger the dispersion. The attenuation in this range increases with frequency in a complicated nonlinear way. The width of the intermediate frequency range decreases significantly as the degree of grain anisotropy increases; 4) At the end of the intermedia range, the dispersion increases dramatically and the velocities of the slow and fast mode quickly approach their high frequency limits. In the high frequency range, the dispersion of both modes become negligible, while the dimensionless attenuation of both modes approaches a constant near unity. We also note several distinct features for longitudinal and transverse waves. For instance, the low frequency branch of the longitudinal waves is always connected to the branch of the fast mode at high frequencies, while the low frequency branch of transverse waves is always connected to the branch of the slow-mode. At high frequencies, the attenuation of the fast-longitudinal mode is always smaller than that of the slow mode. For transverse waves the opposite character is observed.

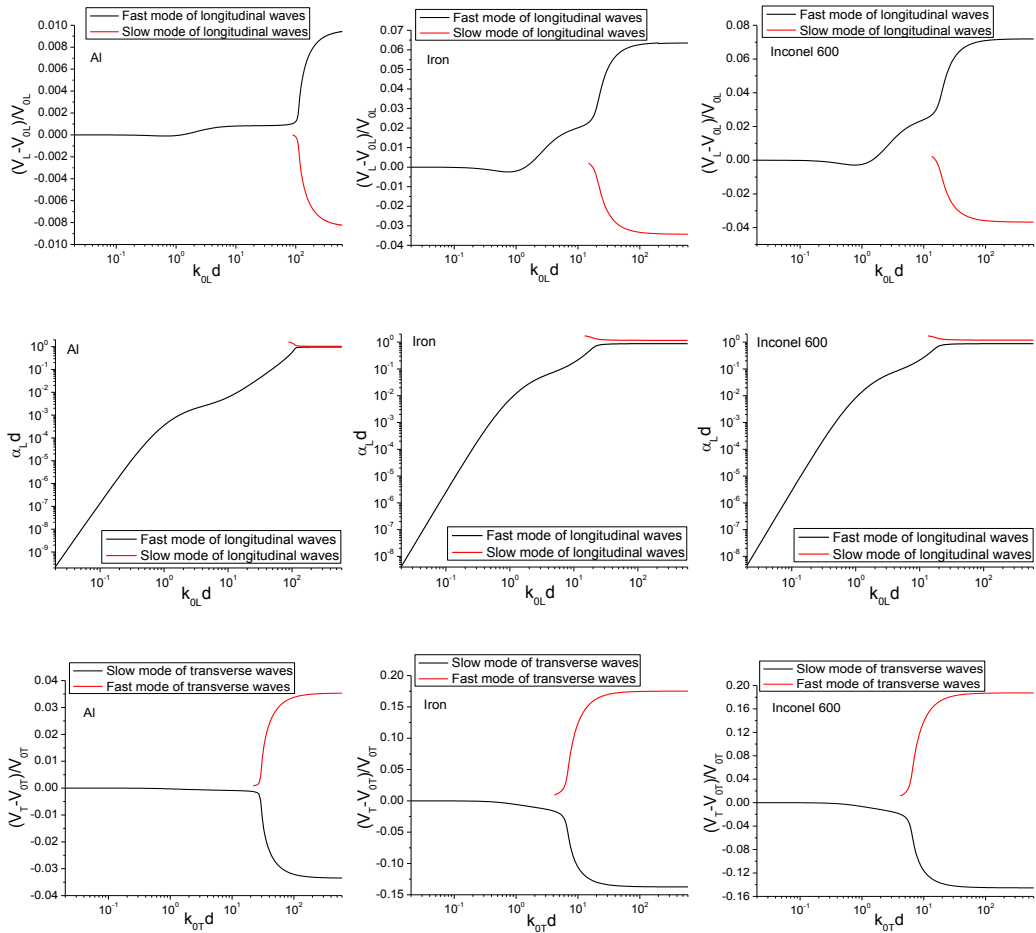

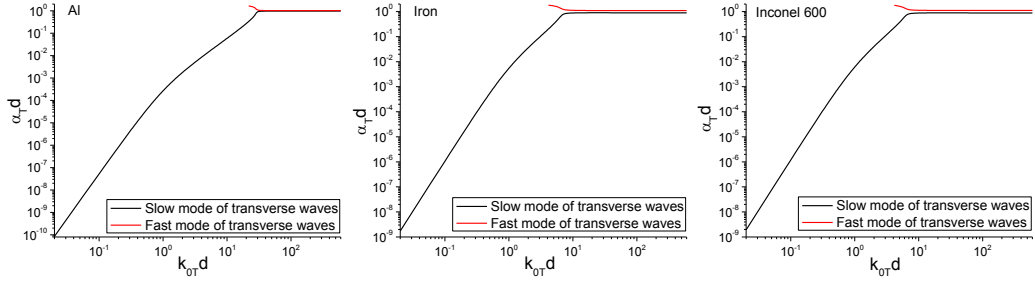

Figure 2 Longitudinal and transverse velocity dispersion and attenuation of

polycrystalline Aluminum, Iron and Inconel 600

Figs. 3 and 4 shows the results obtained by Stanke [31] and that given by the SFMS model for aluminum and iron, respectively. It is observed that throughout the whole frequency range only one propagation mode is obtained by the unified model, while the SFMS model gives two at high frequencies and show the branching phenomenon explicitly. The author believes the unified theory also can predict the other mode but the problem remains open. For aluminum, the predictions given by the unified model are nearly the same as that given by the SFMS theory. Only in the high frequency range $k_0d>100$ can we observe some discrepancy in the dispersion curves. For transverse wave attenuation the prediction of the SFMS theory is slightly larger than that given by the unified theory.

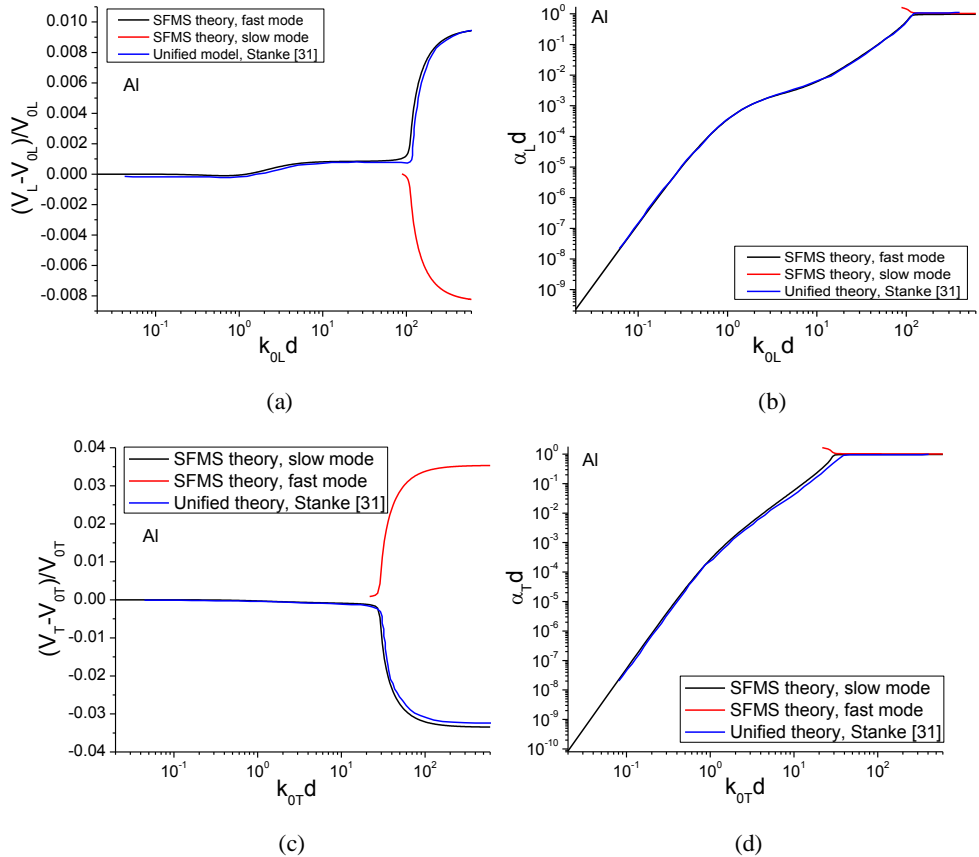

Figure 3 Comparison of the predictions given by the SFMS model and the unified model [31], (a) longitudinal velocity and (b) attenuation, and (c) transverse velocity and (d) attenuation of Al

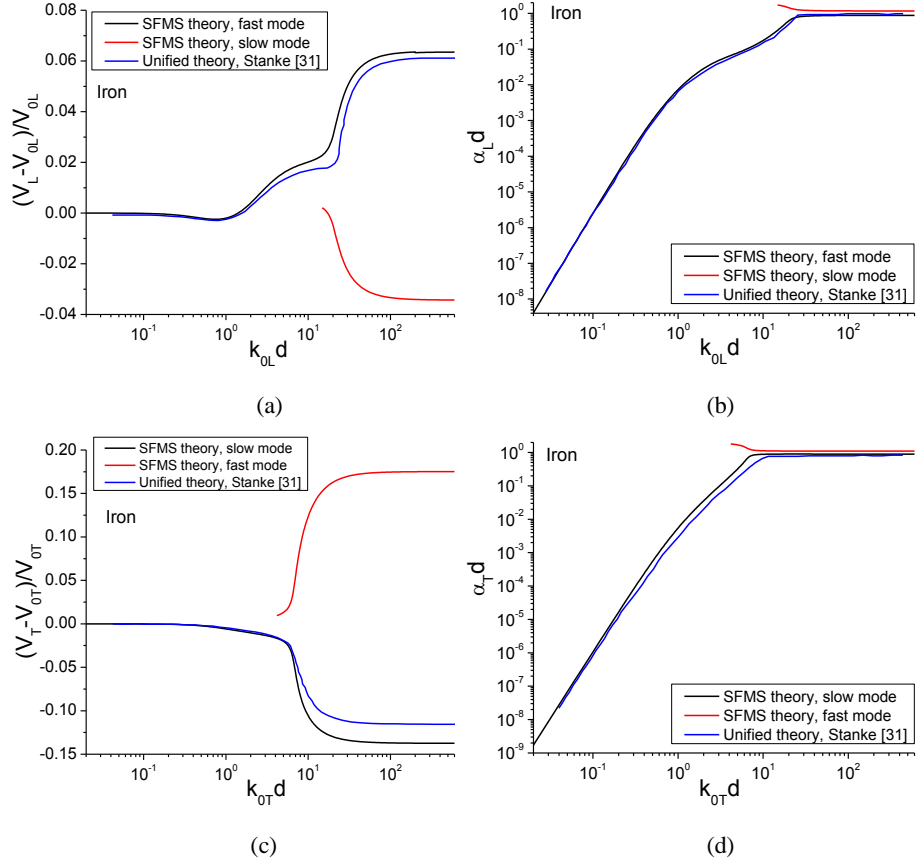

Figure 4 Comparison of the SFMS predictions with that of the unified model [31], (a) longitudinal velocity and (b) attenuation, and (c) transverse velocity and (d) attenuation of iron

Polycrystalline iron is a typical alloy with strong grain anisotropy. The predictions given by the two theories are shown in Fig. 4. For longitudinal velocity and attenuation, the predicted values at low frequency ($k_{0L}d<1$) are nearly identical. At intermediate to high frequencies, the discrepancies between the two models become obvious. The unified theory gives L-wave velocity systematically smaller than that give by the SFMS theory. The maximum value of the discrepancy is about 1%. Meanwhile, the unified theory gives attenuation coefficient systematically lower than that of the SFMS theory. For transverse waves, the two theories give nearly the same results for velocities at low frequencies, $k_{0T}d<1$. At higher frequencies, the predictions given by the unified theory are systematically larger than that of the SFMS theory. The maximum value reaches up to 2% in the geometric regime. The unified theory gives attenuation coefficient uniformly lower than that by the SFMS theory. At intermediate frequencies the predictions given by the former are nearly 50% smaller than the later.

Stanke [31] carried out comprehensive experimental studies on the attenuation behavior of polycrystalline materials. Planer transducers operating in pulse-echo mode and transmission mode are used to measure the attenuation. However, due to the limitation of the measurement system, the phase information of the Fourier transform was missed and as a consequence, the velocity dispersions were not measured. The center frequency of the transducers varies from 10 MHz to 100 MHz, so the attenuation in a relatively broad band, ranging from 16.5 to 85 MHz is obtained. Before comparing the theoretical predictions with that measured in experiments, we first need to discuss several factors that may complicate the measurements and may influence the explanation of the experimental data. The first problem is the proper definition of the spatial autocorrelation distance $a$, as appeared in the expression of the spatial autocorrelation function $\exp(-r/a)$. For an idealized random medium that follows the Poisson statistics strictly, the mean cord length $\bar{c}$ is equal to the half of the mean free path $\bar{d}$ [31]. However, practical polycrystalline materials are not idealized Poisson media, and these two parameters are not always identical. For instance, the ratio for 304 stainless steel varies from 0.74 to 0.81, i.e., $\overline{d}/(2\overline{c}) = 0.74 \sim 0.81$. The second factor worth noting is that the sample may not be a homogeneous material, the grain size varies from point to point, thus the measured signals can be distorted by the variation of grain size. Third, the polycrystal sample may contain twins whose crystallographic orientations are strictly correlated, this factor destroys the assumption that the orientations of different grains are uncorrelated. Fourth, the metal is not pure, actually they are often a complicated system composed of two or more chemical elements. For example, the stainless steel and plain carbon steel can be viewed as a binary system composed of C and Fe. Moreover, the finite beam width of practical transducers inevitably causes distorted signals different from an idealized plane wave signal. The quality of deconvolution method used to eliminate the beam effects also affect the accuracy of final results. With all these complicated factors in mind, Stanke [31] concluded that predictions of the average grain size with less than 20% error can be achieved using the unified theory. Fig. 5 shows the experimental results for attenuation of OFHC copper sample (red line) along with the theoretical prediction given by the SFMS theory. Choosing the experimental data as the standard, the relative error of the theoretical attenuation is less than ±20%. In addition, the overall variation tendency of the experimental curve shows good agreement with the theoretical curves. The accuracy of the SFMS model is fully confirmed by this example.

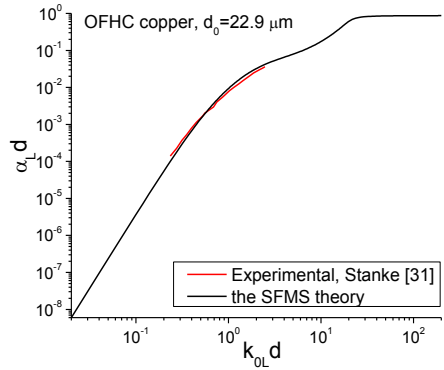

Figure 5 Comparison of the SFMS predictions with experimental data [31] for longitudinal attenuation of OFHC copper

The experimental data for 304 stainless steel are shown in Fig. 6. As pointed out by Stanke [31], the measured steel bar exhibit inhomogeneous microstructures, the grains at the edge of the bar (point d in [31]) is smaller than those at the center (point a). However, the grain size variation is relatively slow and in the scope of the transducer face (11 mm in diameter), the variation can be neglected. The attenuation coefficients measured at point a and d are converted into dimensionless quantities, as marked by the red and blue lines in Fig. 6(a). Once again, the overall variation tendency of the measured attenuation at both points show very good agreement with the theoretical curve. The experimental attenuation normalized using the original grain sizes given by Stanke [31] is slightly smaller than that given by the theory, as discussed above, this may be caused by the inaccurate evaluation of the spatial autocorrelation length. If normalizing the experimental data using 0.9 times of the given grain diameter, as shown in Fig. 6(b), the new set of curves show excellent agreement with that given by the SFMS theory. This example indicates the SFMS theory is capable of predicting the average grain size with relative error less than 10%.

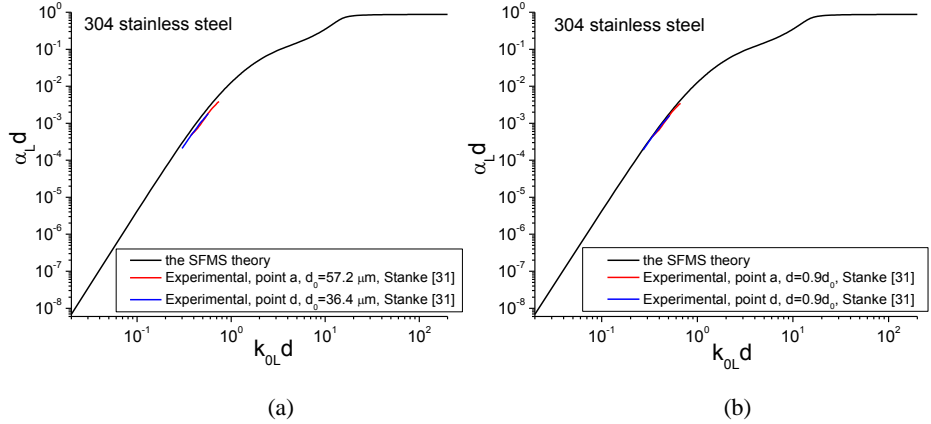

(a)                          (b)

Figure 6 Comparison of the SFMS predictions with experimental data [31] for longitudinal attenuation of 304 stainless steel, (a) normalized using the original average grain diameter, $d_0$= 57.2μm for point a and $d_0$= 36.4μm for point d (b) normalized using new average grain diameters, $d=0.9d_0$

Fig. 7 presents the dispersion and attenuation of an Inconel 600 sample obtained using finite element simulation [57]. The Inconel sample is composed of 5210 randomly oriented grains with an average size of 500 µm. To emulating the pitch-catch measurement of a plane wave signal, a three-cycle tune-burst pulse with a center frequency of 1-3 MHz is applied on one end surface of a prismatic bar, for which the side surfaces are subjected to symmetric boundary conditions. In an idealized case, a large number of realizations of the random media should be generated and the velocity and attenuation of each realization should be calculated to obtain statistically meaningful results. However, due to the high computational cost, only one realization of the material is performed. Fig. 7(a) and (b) demonstrate the longitudinal velocity and attenuation normalized using the original average grain size in [57]. It is seen that the velocity first undergoes slow negative dispersion and then increase with frequency rapidly. This variation tendency is perfectly captured by the SFMS theory. The attenuation obtained using the FEM approach is systematically smaller than that given by the theory. As pointed out by Pamel [57], the predictions of the attenuation is largely dependent on the choice of d. Inspired by this conclusion, we renormalize the experimental data using a slightly smaller grain size, i.e., $d=0.8d_0$, the results are replotted in Fig. 7(c) and (d).

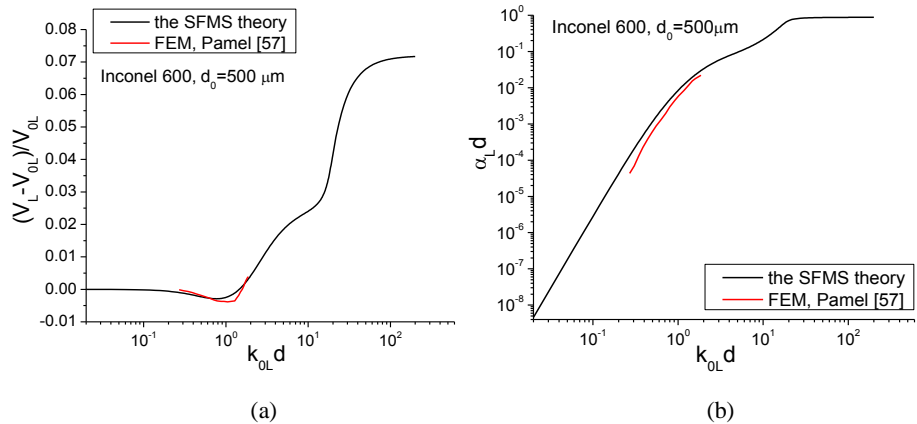

(a)                          (b)

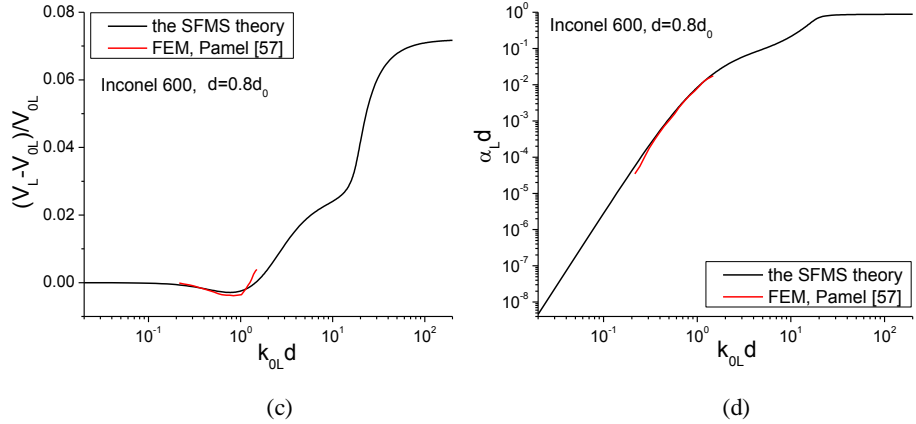

Figure 7 Comparison of the SFMS predictions with FEM simulations [57] (a) longitudinal velocity,
(b) attenuation with original average grain diameter $d_0$=500μm, and (c) longitudinal
velocity, (d) attenuation with best-fitting average grain diameter $d=0.8d_0$

From Fig. 7(c) we see the velocity is still in good agreement with that predicted by the theory. Surprisingly, the dimensionless attenuation given by the numerical results show excellent agreement with the theoretical curves, as shown in Fig. 7(d) gives. This example shows that the SFMS theory is capable of providing a satisfactory prediction for the velocity and attenuation of coherent waves in strongly scattering polycrystals. Since the numerical algorithm and the statistics of the grains have certain errors intrinsically, we do not give an estimate on the relative error.

Through the above examples we see the SFMS theory is capable of providing accurate predictions for both the velocity and attenuation of a large variety of polycrystalline materials. Comparison with the unified theory reveals that for polycrystalline materials with weakly anisotropic grains, the two theories provide nearly the same predictions, but for materials with strongly anisotropic grains, there are noticeable discrepancies between the two models. The unified theory gives smaller longitudinal velocity and larger transverse velocity at intermediate or high frequencies. It also gives attenuation coefficient systematically smaller than that given by the SFMS theory. However, more accurate numerical simulations and experimental studies are in need to justify which model is more accurate, especially in the high frequency range. Comparison with experimental results further confirm the accuracy of the SFMS theory, for most materials the theoretical predictions show very good agreement with experimental data. The results also suggest that if the basic assumptions are satisfied, i.e., statistically homogeneous, untextured with equiaxed grains, the SFMS theory is capable of predicting the average grain size with a relative error less than 10%.

## IV. PRACTICAL APPLICATIONS OF THE NEW MODEL

In this section, we demonstrate the practical applications of the new model in nondestructive characterization of microstructures in the most frequently used jet engine alloys Ti64 and Ti6242. The velocity and attenuation of longitudinal and transverse coherent waves for pure titanium and its alloys are calculated. Both alpha (with hexagonal close-packed grains, HCP) and beta (with body-centered cubic grains, BCC) phases are considered. Special attentions are paid to the sensitivity of ultrasonic propagation parameters to the interested microstructural characteristics, like grain size, crystallite symmetry, and chemical composition of alloys. As applications in seismology, we calculate the velocity and Q-factors of seismic waves in realistic iron models of the Earth's uppermost inner core. The results are used to explain the observed velocity and attenuation. Most importantly, the consistent between the theoretical predictions and the measured data poses fundamental constraints on the correlation length (size of macrograins) and the grain anisotropy of the inner core.

**A) Ultrasonic nondestructive characterization of microstructures in Titanium alloys**

Pure titanium exists in the form of alpha phase under 882.5 °C, in the range of temperature between 882.5 °C to 1668 °C, it exists in the form of beta phase. With the addition of alpha stabilizer or beta stabilizer, both the alpha and the beta phase can exist in a broad range of temperature, thus the titanium alloys in the form of alpha, beta or alpha+beta phases are all used [1-3]. Titanium alloys constitute the most important jet engine materials. Nondestructively evaluating the grain size, volumetric ratio of the alpha to beta phases, and monitoring the elastic property changes are the major purpose of ultrasonic nondestructive technologies [10, 15, 59-61]. As the first step towards the development of a comprehensive theoretical system for modeling ultrasonic scattering in engine-grade titanium alloys with complicated microstructures, we first calculate the dispersion and attenuation of pure Ti, Ti64 and Ti6242. For each type of alloys, we consider both the (near) α phase and the (near) β phase. The material parameters used in this work are shown in Tab. 4. Tab. 5 gives the reference properties and the Voigt average properties.

Table 4 Material properties of Titanium and its alloys

| Material | | Elastic constants (GPa) | | | | | | Density |
|---|---|---|---|---|---|---|---|---|
| | | $C_{11}$ | $C_{12}$ | $C_{13}$ | $C_{33}$ | $C_{44}$ | $C_{66}$ | $\rho$ (kg/m$^3$) |
| Pure α-Ti [15] | HCP | 162.0 | 92.0 | 69.0 | 180.0 | 46.7 | 35.0 | 4540 |
| Pure β-Ti (1020 °C) [15] | BCC | 129 | 101 | 101 | 129 | 37 | 37 | 4500 |
| α-Ti64 [61] | HCP | 174.4 | 98.1 | 72.0 | 197.3 | 50.7 | 38.15 | 4540 |
| β-Ti64 [61] | BCC | 151.2 | 108.0 | 108.0 | 151.2 | 41.1 | 41.1 | 4480 |
| α-Ti6242 [62] | HCP | 170.0 | 98.0 | 86.0 | 204.0 | 51.0 | 36.0 | 4540 |
| β-Ti6242 [62] | BCC | 250.2 | 119.0 | 119.0 | 250.2 | 115.3 | 115.3 | 4540 |

Table 5 Reference velocities of Titanium and its alloys (Unit: m/s)

| Material | Reference property | | Voigt property | | Reference velocity | | Voigt velocity | |
|---|---|---|---|---|---|---|---|---|
| | $\lambda$ (GPa) | $\mu$ (GPa) | $\bar{\lambda}$ (GPa) | $\bar{\mu}$ (GPa) | $V_{0L}$ | $V_{0T}$ | $\bar{V}_L$ | $\bar{V}_T$ |
| Pure α-Ti | 78.29 | 43.29 | 77.81 | 43.95 | 6026.19 | 3087.92 | 6041.52 | 3111.37 |
| Pure β-Ti | 80.81 | 21.95 | 79.68 | 23.48 | 5264.35 | 2208.57 | 5304.92 | 2284.25 |
| α-Ti64 | 82.93 | 47.39 | 82.36 | 48.18 | 6256.45 | 3230.84 | 6274.20 | 3257.66 |
| β-Ti64 | 101.08 | 32.11 | 100.2 | 33.3 | 6074.32 | 2677.20 | 6101.81 | 2726.36 |
| α-Ti6242 | 90.24 | 45.20 | 89.87 | 45.87 | 6307.82 | 3155.30 | 6324.73 | 3178.60 |
| β-Ti6242 | 101.26 | 92.53 | 99.12 | 95.42 | 7941.42 | 4514.54 | 7991.74 | 4584.50 |

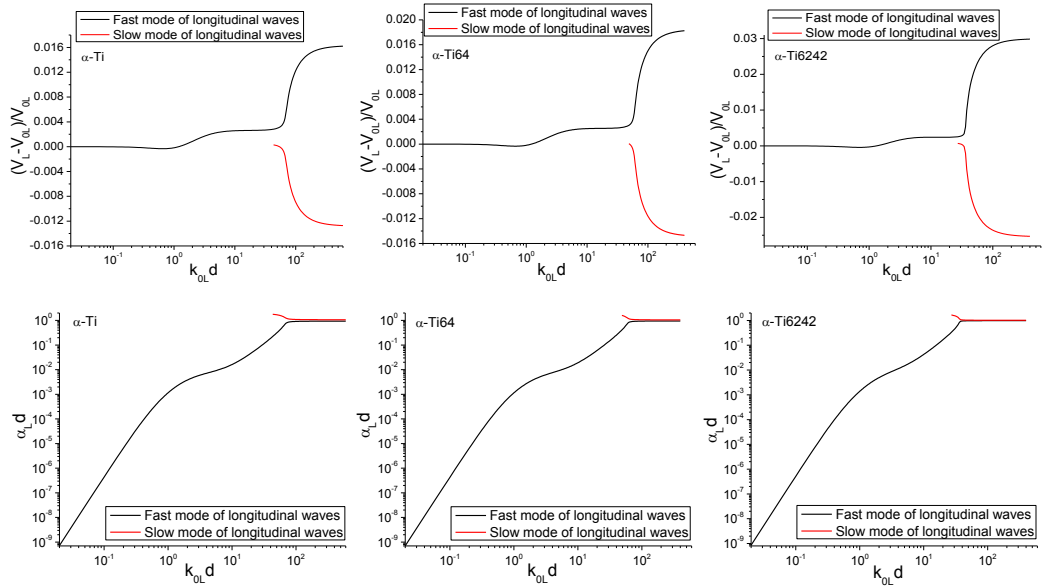

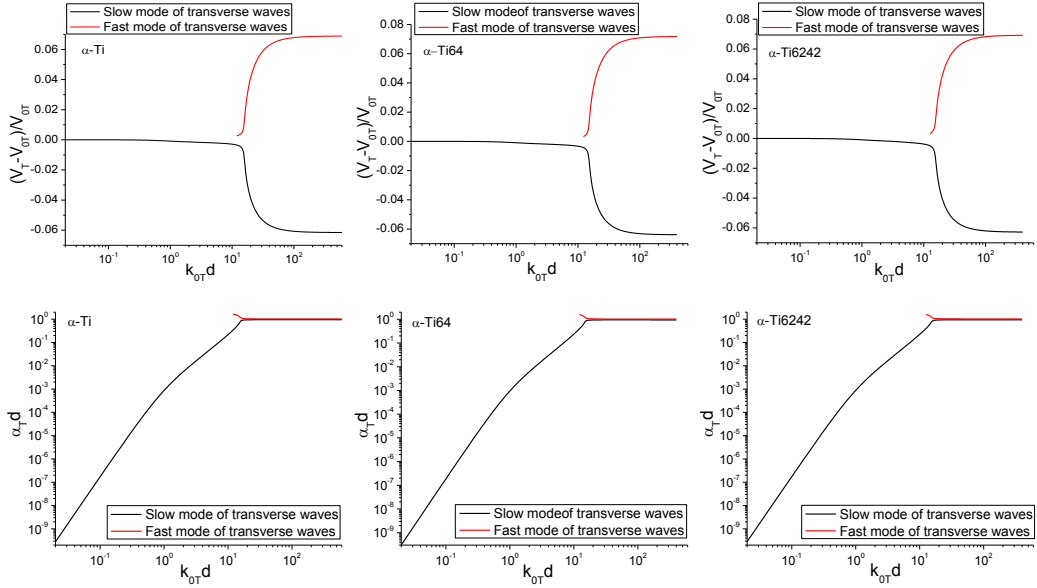

Figure 8 Velocity and attenuation of alpha phase titanium and its alloys

Figure 8 shows the velocity and attenuation of L and T waves in alpha phase Titanium alloys. Each dispersion curve starts from the wave velocity of the reference media (the quasi-static limit of the coherent wave). As pointed out before, one of the major difference between the unified theory and the SFMS theory is that they use different reference media: the former chooses the Voigt average properties for the homogeneous reference medium, while the later selects the properties of the reference medium by enforcing the volumetric average of the renormalized property fluctuation $\varXi_{ab}(\mathbf{x})$ vanish. This is one of the most important strategies adopted by the new theory to secure the fastest rate of convergence. In the low frequency regime $k_{0L}d<1$, the dispersion of the velocity is nearly negligible. In the intermediate frequency range, the velocity increases to a value slightly larger than the quasi-static limit, and then become constant again. This is a common characteristic of weak property fluctuation media, as that observed in the dispersion curves of Al alloy in Fig. 2. In the transition regime from intermediate to high frequencies, the velocity increase rapidly to its geometric limit, meanwhile, a second, slow mode starts to appear and its velocity decreases rapidly to the lower geometric limit. The difference between the upper and lower geometric limits is closely related to the fluctuation of material properties, i.e., the degree of grain anisotropy. The relatively small difference in these two limits indicates the alpha phase titanium possess relatively weak grain anisotropy. The variation tendency of attenuation coefficients of the alloys is similar to that for the materials in Fig. 2. One unique feature of the L-wave attenuations is that they all exhibit a hump in the intermediate region $1<k_{0L}d<10$, which is caused by the mode conversion from longitudinal to transverse waves. The dispersion curves of transverse waves have a relatively simple pattern. The velocity remains constant in the broad band $0<k_{0T}d<10$. At the end of the intermediate regime, a second, fast mode starts to appear and then the velocity of both the modes rapidly approaches the higher and lower geometric limits.

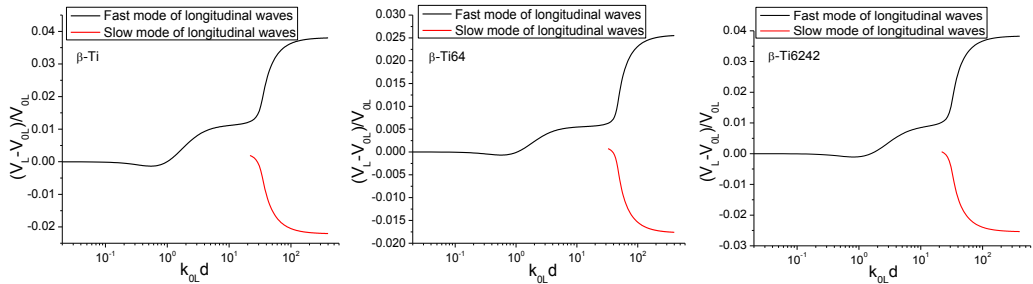

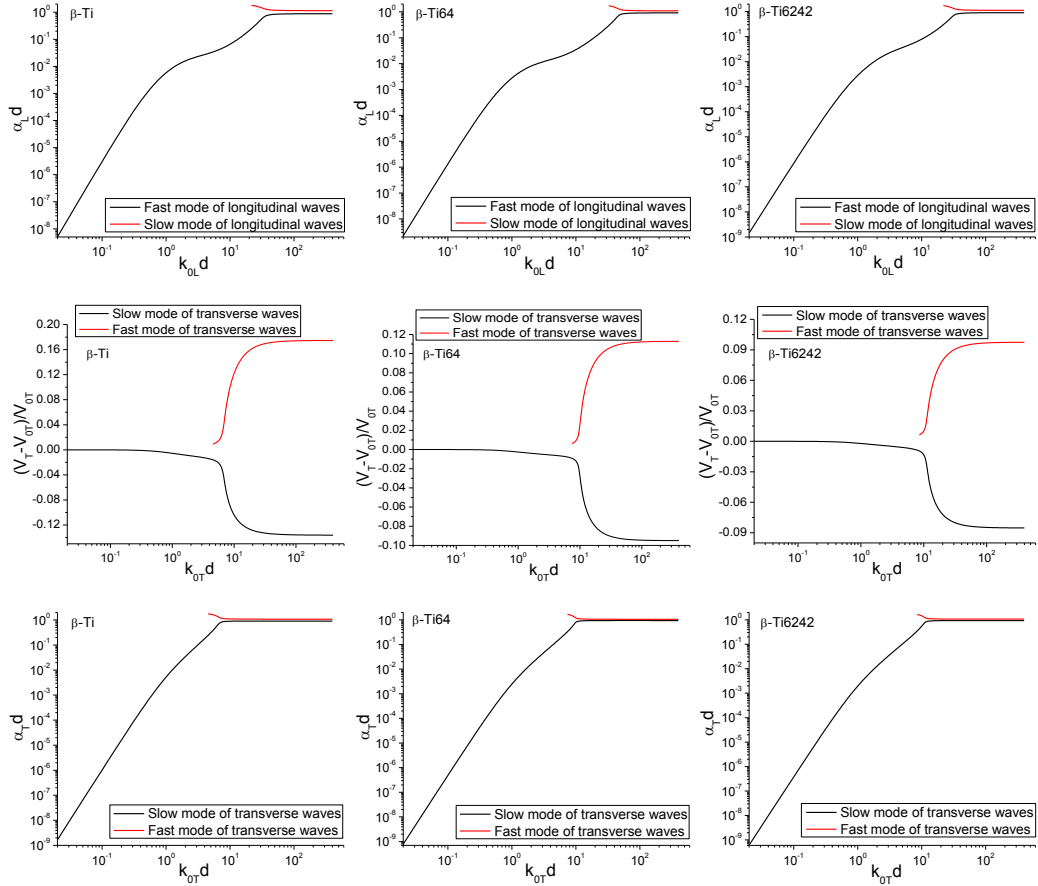

Figure 9 Velocity and attenuation of beta phase titanium and its alloys

The velocity and attenuation of beta phase Titanium alloys are shown in Fig. 9. The overall variation tendencies of the velocity and attenuation are similar to those of the alpha phase. It is seen the beta phase alloys have relatively strong grain anisotropy, so the difference between the two modes at geometric regime becomes large, meanwhile, the longitudinal velocity undergoes continuous positive dispersion in the intermediate range, where the "plateau stage" for weak fluctuation alloys nearly disappears. The grain anisotropy has more obvious effects on the transverse velocities which lead to an exceedingly large difference between the two geometric limits, for instance, the upper geometric limit of β-Ti raises by 18% and its lower limit decreases by -13% relative to the reference velocity.

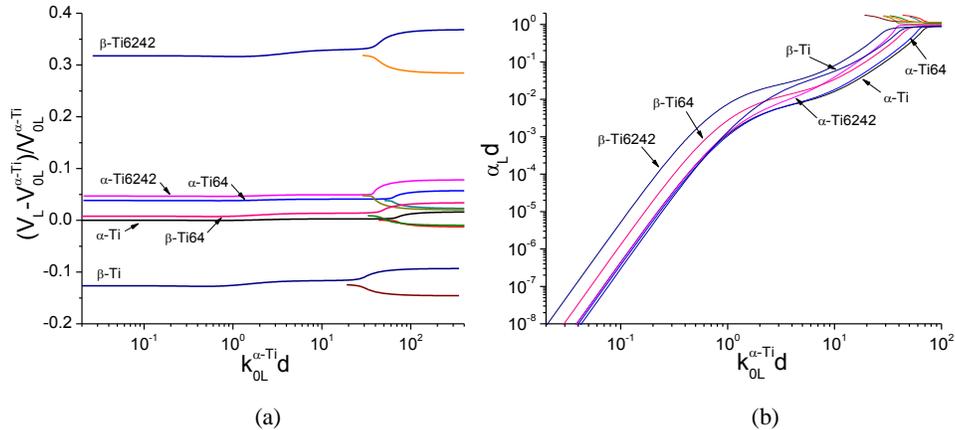

(a)   (b)

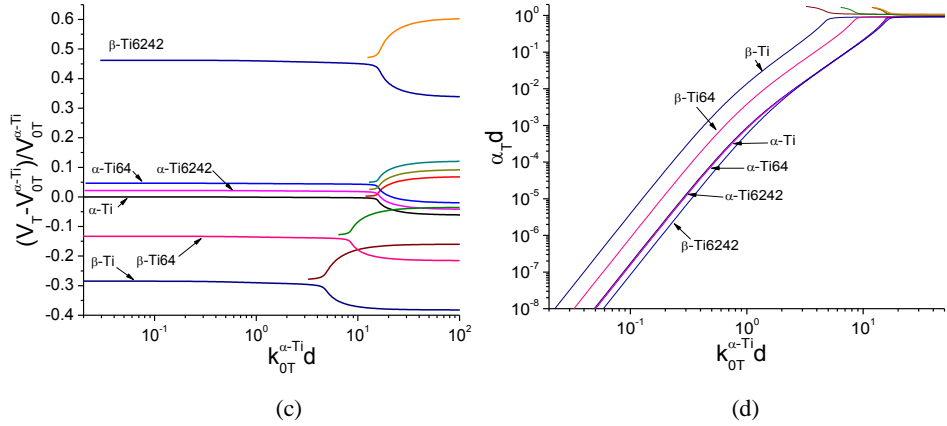

(c)             (d)

Figure 10 Sensitivity of velocity and attenuation to the phases and alloying elements of Titanium alloys:

(a) longitudinal velocity, (b) longitudinal attenuation, (c) transverse velocity, (d) transverse attenuation

To facilitate a comparison between the alpha and beta phase alloys and to observe the different variation tendencies of different alloys, the dispersion and attenuation curves are plotted in the same figure, see Fig. 10, where all the quantities are normalized to those of the α-Ti. As can be seen, for the same alloy the dispersion and attenuation curves of the alpha and beta phases show clear distinctions. This important feature has significant implications for monitoring the manufacturing process of titanium alloys. The alloy billet or repaired parts normally undergoes a series of complicated heat treatments, including heating, annealing, quenching, during which the microstructures experienced complex transitions between alpha phase and beta phase. Ultrasonic monitoring the phase transition during these processing procedures helps us gain a better understanding of the microstructure evolution, and thus provides important technical strategy to ensure that the desired phase or phase composition are achieved.

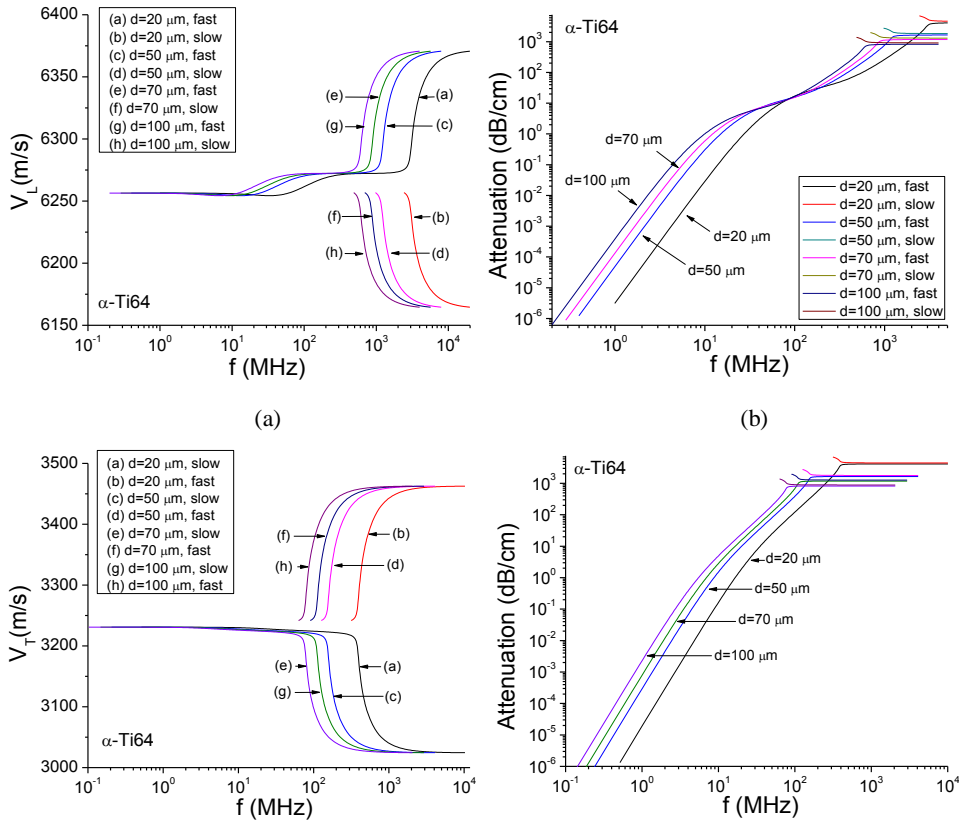

(c)                                                                    (d)

Figure 11 Effects of grain size on the velocity and attenuation of Titanium alloy α-Ti64

Ultrasonic dispersion and attenuation provide two important metrics for quantitative measurement of grain sizes [63-64]. We take α-Ti64 as an example to show the grain size effects on the dispersion and attenuation of ultrasonic waves. The dispersion and attenuation curves for α-Ti64 with different average grain size, varying from 20 μm to 100 μm, are depicted in Fig. 11. It is seen the dispersion and attenuation curves follow a scaling law, i.e., with the increase of grain diameter, the dispersion curves are scaled down along the frequency axis and the attenuation curves are scaled down along both axes. The longitudinal velocities at low frequencies nearly independent of the grain size. In the intermediate frequency range ($10 < f < 200$ MHz), the dispersion curves show obvious dependence on the grain size. The opposite behavior is observed for the size dependence of the attenuation curves. At low frequencies ($0 < f < 20$ MHz), the attenuation is strongly dependent on the grain size, the larger the grain diameter, the larger the attenuation. In the intermediate range ($20 < f < 200$ MHz), the dependence of attenuation on the grain size become weaker, which indicating that the attenuation of low frequency signals and the dispersion of intermediate frequency signals are good metrics of the average grain size. In the geometric regime ($f > 500$ MHz), the attenuation is independent on the frequency and inversely proportional to the average grain diameter. However, the very high frequency signals are rarely used in practical measurements due to its high attenuation. For transverse waves, the velocity nearly keeps constant in the broad frequency region ($0 < f < 70$ MHz), so it is not a promising candidate for the measurement of grain size. Contrarily, the attenuation in this frequency regime shows strong dependence on the grain size, which indicates the transverse attenuation is a good indicator of the grain size.

In summary, the full-frequency range velocity and attenuation for the most frequently used jet engine alloys are obtained for the first time. Through comprehensive study on the effects of alloy phases and grain sizes on the ultrasonic propagation parameters, we demonstrate the great potential of the new multiple scattering theory in applications in quantitative characterization of microstructures in jet engine alloys. The new model is of great importance for the development of the next generation ultrasonic nondestructive inspection techniques for jet engine manufacturing and maintenance.

**B) velocity and attenuation of seismic waves in the Earth's uppermost inner core**

Seismic waves carry rich information about the structures and properties of the traversed volumes. Its unique capability of penetrating an exceedingly large depth into the Earth makes it one of the most important methods to detecting the structure of Earth's inner core. The material composition and small-scale structure of the Earth's inner core has drawn extensive attention from the geophysical and geochemical communities. Since direct measurement is prohibited by the large depth of the inner core, researchers can only conduct numerical simulations [65-67] or laboratory measurements [68] based on primary knowledge of the high temperature and high-pressure conditions that exist in the inner core. As discussed in [25], the inner core is most likely composed of polycrystalline iron or its alloys. Geochemical studies suggest that the iron crystallites may exist in two forms, i.e., the hexagonal close-packed (HCP) structure and the body-centered cubic (BCC) structure, under the extreme conditions at the inner core. First principle simulation has been proved a powerful tool for deriving the macroscopic elastic properties from the interatomic potentials when the external temperature and pressure conditions are specified. Using the obtained elastic properties, quantitative seismic scattering models can predict the accurate velocity and attenuation behavior of the material model, which in turn can be used to validate and improve the proposed material model through comparison with the measured seismic data. In the following we calculate the dispersion and Q-factors of six polycrystalline iron models, and try to give an explanation to the observed velocity and attenuation in the scattering scenario. The material properties used are adopted from Laio et. al. [65], Vocadlo [66], Belonoshko et. al. [67], and Mao et. al. [68], as listed in Tab. 6.

Table 6. Material properties of iron polycrystalline models of the Earth uppermost inner core

| Material model | | Elastic constants (GPa) | | | | | | Density |
|---|---|---|---|---|---|---|---|---|
| | | $C_{11}$ | $C_{12}$ | $C_{13}$ | $C_{33}$ | $C_{44}$ | $C_{66}$ | $\rho$ (kg/m$^3$) |
| 1 [68] | HCP | 1533 | 846 | 835 | 1544 | 583 | 343.5 | 12610 |

| | | | | | | | |
|---|---|---|---|---|---|---|---|
| 2 [65] | HCP | 1697 | 809 | 757 | 1799 | 421 | 444 | 12885 |
| 3 [66] | HCP | 1730 | 1311 | 1074 | 1642 | 159 | 209.5 | 13155 |
| 4 [67] | BCC | 1561.6 | 1448.1 | 1448.1 | 1561.6 | 365.5 | 365.5 | 13580 |
| 5 [66] | BCC | 1603 | 1258 | 1258 | 1603 | 256 | 256 | 13155 |
| 6 [66] | BCC | 1795 | 1519 | 1519 | 1795 | 323 | 323 | 13842 |

The properties of the homogeneous reference media and the Voigt average properties are listed in Tab. 7. It is seen the velocities of these material are much higher than those under normal conditions due to the action of extremely high pressure and very high temperature.

Table 7. Reference velocities of iron polycrystalline models of the Earth uppermost inner core

| Material model | Reference property | | Voigt property | | Reference velocity | | Voigt velocity (km/s) | |
|---|---|---|---|---|---|---|---|---|
| | $\lambda$ (GPa) | $\mu$ (GPa) | $\bar{\lambda}$ (GPa) | $\bar{\mu}$ (GPa) | $V_{0L}$ (m/s) | $V_{0T}$ (m/s) | $\bar{V}_L$ (m/s) | $\bar{V}_T$ (m/s) |
| 1 | 786.2 | 428.9 | 777.0 | 441.5 | 11418.09 | 5832.04 | 11473.52 | 5917.09 |
| 2 | 794.9 | 447.5 | 794.2 | 448.5 | 11452.18 | 5893.24 | 11456.59 | 5899.82 |
| 3 | 1195.3 | 207.3 | 1192.2 | 215.0 | 11062.52 | 3969.67 | 11104.70 | 4042.72 |
| 4 | 1358.6 | 195.8 | 1324.6 | 242.0 | 11352.56 | 3797.14 | 11540.41 | 4221.41 |
| 5 | 1226.9 | 219.5 | 1224.6 | 222.6 | 11253.28 | 4084.81 | 11266.44 | 4113.55 |
| 6 | 1455.4 | 234.9 | 1445.0 | 249.0 | 11793.39 | 4119.48 | 11847.78 | 4241.31 |

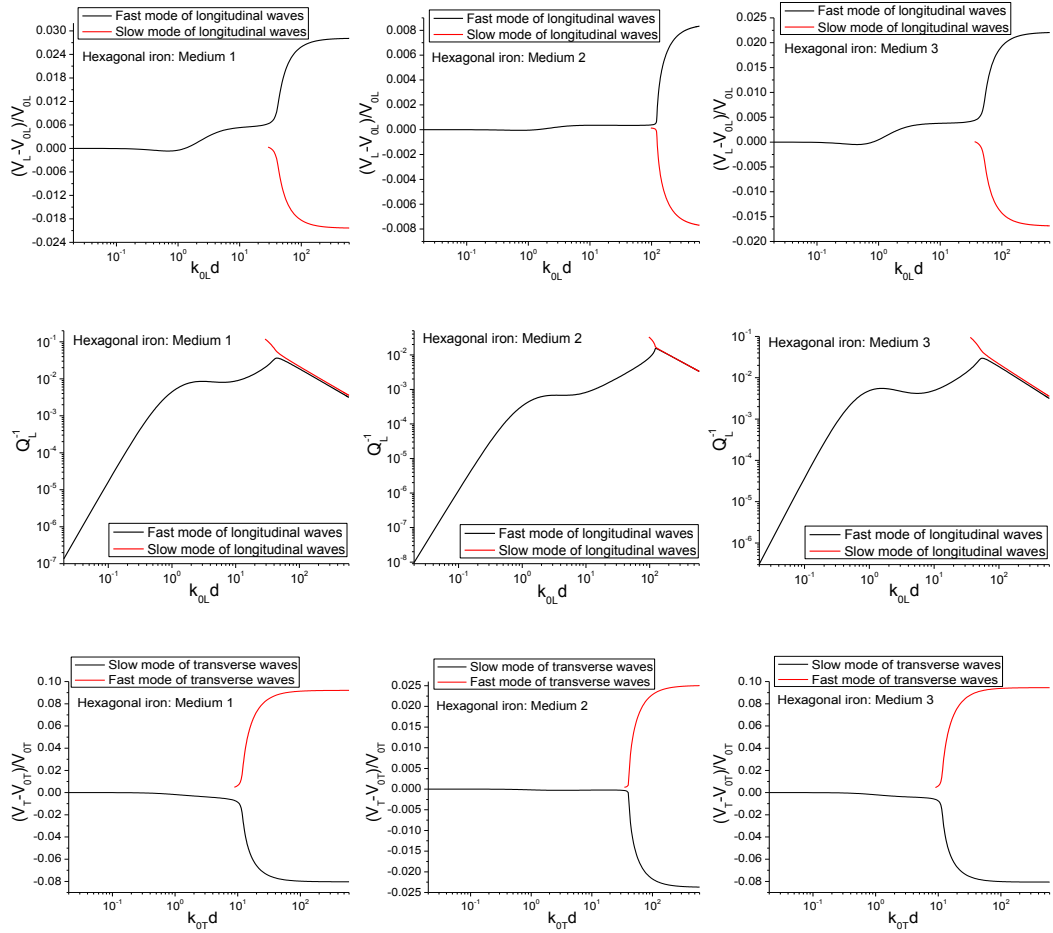

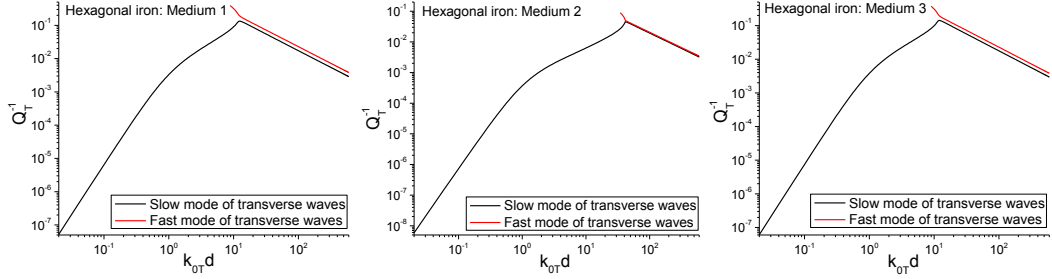

Figure 12 Velocity and Q-factors of hexagonal iron models of the Earth's uppermost inner core

The dispersion and Q-factors of longitudinal and transverse waves of hexagonal iron models are presented in Fig. 12. The Q-factors (actually the inverse Q-factors) are calculated by $Q^{-1} = \mathrm{Im}(k^2)/\mathrm{Re}(k^2)$, as elaborated in [51]. It describes the attenuation of the coherent waves. Similar to the examples discussed above, there is only one propagation mode at low and intermediate frequencies, and two modes at high frequencies. The longitudinal velocities in the high frequency regime only deviate slightly from those of the reference media, from ±0.8% for Medium 2 to ±2.5% for Medium 1. However, the velocities of transverse waves in the high frequency regime exhibit very large variations, reaching up to ±10% for Media 1 and 3. The Q-factors of longitudinal waves demonstrate a very interesting variation tendency. At low frequencies ($0<k_{0L}d<1$) it increases with frequency following a power law, then the slope decreases rapidly and the Q-factor assumes a local maximum. After that the Q-factors decrease slightly and reach a local minimum at $k_{0L}d\approx10$. As the frequency increasing, the Q-factor increases again and reaches its maximum value at $k_{0L}d=40\sim100$. In the geometric regime, the Q-factors of both modes decrease with frequency following an inverse power law.

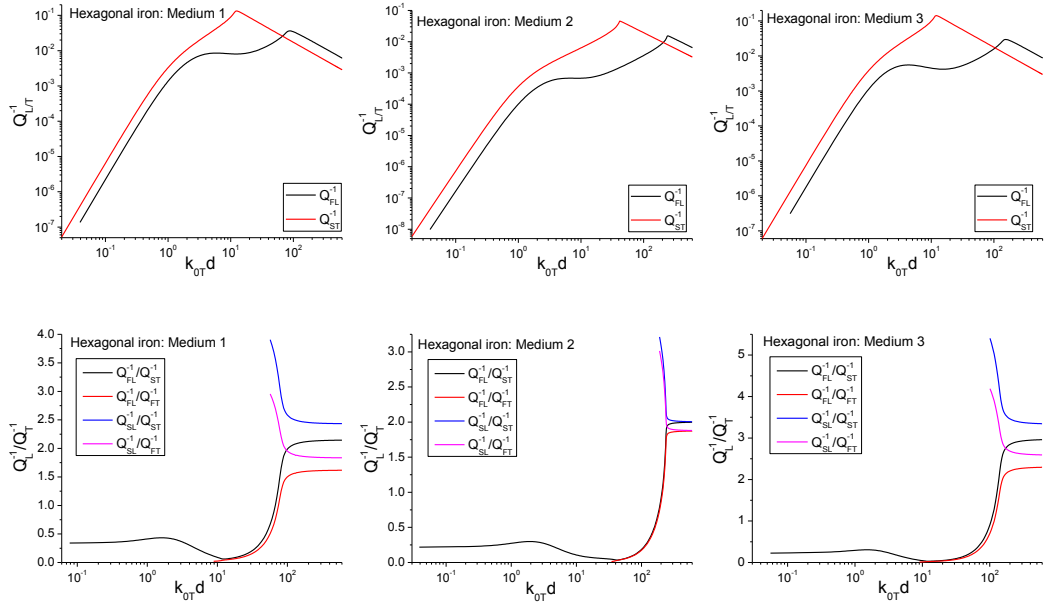

Figure 13 Q-factors and Q-factor ratios of hexagonal iron models of the Earth's uppermost inner core

To compare their relative magnitudes, the longitudinal and transverse Q-factors are plotted in the same frequency scale, as shown in the first row of Fig. 13. From the results we can see from low to intermediate frequency, the transverse Q-factors are systematically larger than the longitudinal Q-factors, but in the high frequency region, the relative magnitude reverses, which tells us at low to intermediate frequencies the attenuation of transverse waves is larger than that of longitudinal waves, while in the high frequency regime, the attenuation of transverse waves become smaller than that of longitudinal waves. The ratios of the longitudinal to transverse Q-factors are depicted in the second row in Fig. 13. The results for Q-factor ratios further tell us the Q-factor ratios in the frequency range $0<k_{0T}d<60$ is very small, normally well below 0.5. In the high frequency region, the four ratios approach four different geometric limits, ranging from 1.5 to 3.5.

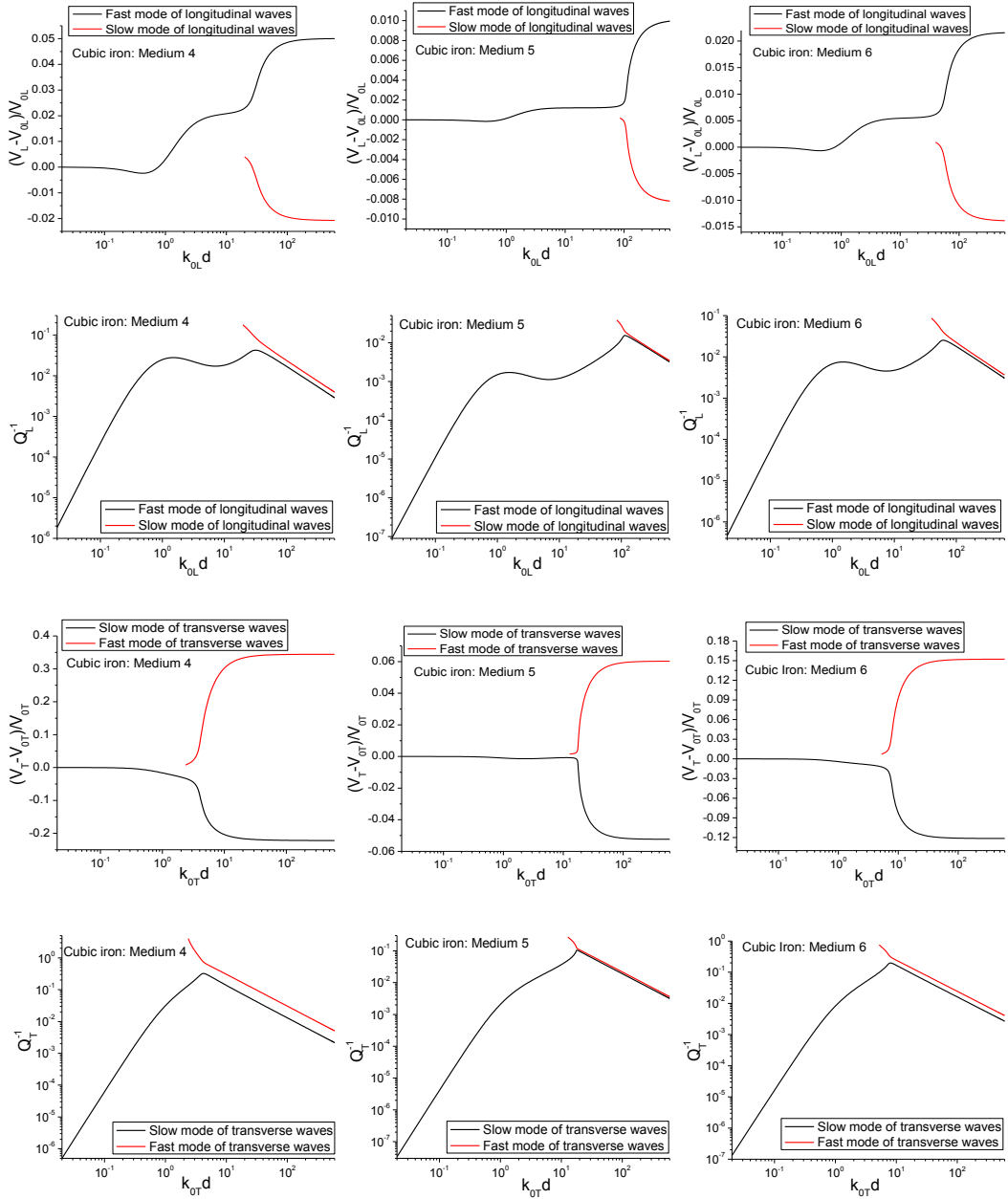

Figure 14 Velocity and Q-factors of cubic iron models of the Earth's uppermost inner core

The dispersion and Q-factors for cubic iron mode are shown in Fig. 14. The general features are similar to those of hexagonal iron models. One noticeable characteristic is that the two maxima and one minimum are more obvious than those of hexagonal iron. The geometric limits of the two transverse modes of Medium 4 deviate significantly from the reference medium for which the lower limit is 20% less than the reference velocity.

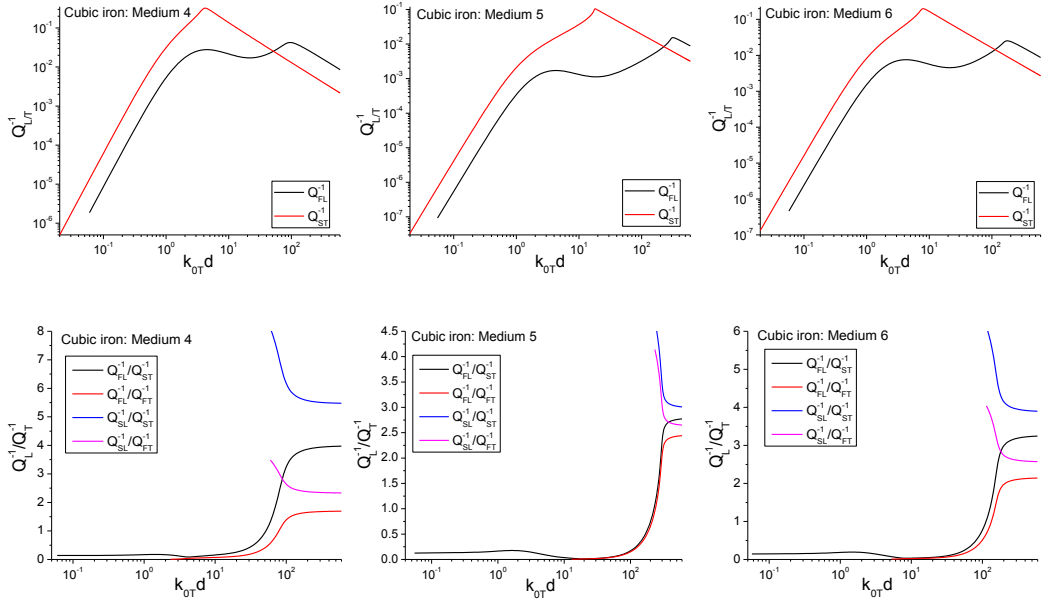

Figure 15 Q-factors and Q-factor ratios of cubic iron models of the Earth's uppermost inner core

The Q-factor ratios of cubic iron models are shown in Fig. 15. The remarkable feature is that at relatively low frequencies the value of the ratios become exceedingly small, uniformly less than 0.2, indicating that the attenuation of longitudinal waves in this range is much less than that of transverse waves. As a result of the large grain anisotropy of Medium 4, the geometric limits of Q-factor ratios vary in a broader range, from 1.5 to 5.5.

The above numerical examples provide us a more explicit picture about the possible dispersion and attenuation of seismic waves propagating in the polycrystalline inner core. With these valuable knowledge, now we try to give an explanation to a series of longstanding problems which are actively discussed in the seismological community. Seismic waves reflected form the inner core like PKiKP waves exhibit strong attenuation. The inverse Q-factors of P waves varies between 0.0025 to 0.01. Meanwhile, the reflected P waves are also accompanied by strong coda, which are strong evidence indicating that the scattering attenuation is the dominant mechanism for the seismic attenuation. The observed P wave velocity (around 11000 m/s) shows good agreement with the Voigt average velocity of the material models, the difference between the two is normally less than 2%. However, the observed S wave velocity (3.5 to 3.67 km/s) is much smaller than the Voigt velocity (4.0 to 4.4 km/s) of the proposed material models. The discrepancy can be as large as 10% to 15%. This has become a longstanding issue in geophysics. Some research proposed that there must be some degree of melting in the inner core. Other researchers believe that the low transverse velocity is caused by impurities or defects. Here we propose an explanation based on scattering. It is generally accepted that the average grain size lies between 200 m to 90 km. From Tab. 7 we can see the Voigt velocity of the longitudinal (P) waves are nearly the same as that of the reference velocity used in this work. In the following analysis we do not discriminate these two velocities and called both of them the Voigt velocity. The Voigt velocity of transverse (S) waves are also very close to the reference velocities used in this work except for Medium 4, so we do not distinguish these two quantities either. For simplicity, we choose the value of the Voigt velocity as: $V_L^{Voigt} = 11000$ m/s, $V_T^{Voigt} = 4000$ m/s. In order to evaluated the velocity and Q-factors, we first need to estimate the dimensionless wavenumber $k_{0L}d$ and $k_{0T}d$. Since the typical frequency of short period PKiKP waves is around 1 Hz, we adopt $f$=1 Hz. If using the relatively small diameter d=200m, then we get $k_{0L}d$=0.11, $k_{0T}d$=0.314. If using the relatively large diameter, d=90 km, then we have $k_{0L}d$=51.38, $k_{0T}d$=141.3. For polycrystal model with small grain size, for example, d=200 m, the velocity of both P and S waves are nearly the same as the Voigt velocities, this is in contradiction with the fact that the velocities of the S waves are normally 10% less than the Voigt velocity. Meanwhile, the Q-factors of P waves lie between $10^{-6}$ to $10^{-4}$, much less than the observed values, normally between 0.0025 to 0.01. All these facts clearly indicate the average grain size is larger than 200m. For polycrystal model with relatively large grain size, i.e., d=90 km, we can see from Figs. 12 and 14 the P wave velocities of all the material models are still close to the Voigt velocity, varying within 2% relative to the Voigt average velocity. However, the S wave velocity of the slow mode, which is also the major mode with smaller attenuation and carrying more energy, is much smaller than the Voigt average, varying from 8% for Media 1, 3 to 20% for Medium 4. Meanwhile, the Q-factors of the P waves lie between 0.001 and 0.01, which shows good agreement with the measured data. From the above analysis we can see the new model is able to give a consistent explanation to the observed phenomena, including the abnormally slow S wave velocity and the Q-factors of P waves. The agreement between the predicted Q-factors with that measured in practical seismic data strongly shows that scattering plays a central role in determining the seismic wave attenuation. In the new model we neither introduce the melting mechanism to explain the S wave velocity, nor do we introduce viscoelastic mechanism to explain the attenuation. Finally, we need to point out that with the development of seismic imaging and signal processing technologies, along with more and more detailed knowledge of the structure and scattering behavior of the mantle and the lithosphere, it is possible to eliminate the influence of the mantle and the lithosphere heterogeneities, and extract the dispersion and Q-factors that exactly corresponds to the inner core. With the accurate experimental data, the grain anisotropy and the average grain size of the inner core can be inversely determined with the new model.

## V. DISCUSSION

The comprehensive numerical study reveals several common features of the coherent waves in various polycrystals. At low frequencies, both longitudinal and transverse waves have a single mode with a velocity near that of the reference medium. In the high frequency regime there are two propagation modes for both the longitudinal and transverse waves. The dispersion of the waves at low and high frequencies is very small, indicating the wave packet can propagate a long distance. In the intermediate frequency range, the longitudinal velocity slightly increases while the transverse velocity gradually decreases. The attenuation coefficients of both the longitudinal and the transverse waves increase following a power law at low frequency, and become constant near unity at high frequency. In the intermediate frequency range the longitudinal attenuation exhibits a hump, indicating strong mode conversion occurs. The Q-factors of both longitudinal and transverse waves increase following a power law at low frequency and decrease following an inverse power law at high frequency. The most interesting phenomenon is the longitudinal Q-factors show an additional maximum in the intermediate frequency range. Comparing the results in this work with those in the previous work [51], we see the bifurcation of the dispersion curves is a universal property of coherent waves in heterogeneous media, regardless of polycrystals or two-phase materials. The discoveries made in this work are of major importance to ultrasonic applications and seismology. The average grain size of common polycrystals used in industry lies between several micros to a hundred micro, and the center frequency of most transducers varies from 0.5 MHz to 100 MHz, thus the ultrasonic waves actually lie in the low frequency regime in the scattering sense, i.e., $k_0d<10$. Consequently the two propagation modes are rarely observed in the ultrasonic NDE. In contrary to industrial materials, the inhomogeneities in the Earth have a rather large characteristic size, normally lies between several kilometers to tens of kilometers, meanwhile, the frequency of seismic waves covers a broad band, from 0.01 Hz to 100 Hz, thus the dimensionless frequency $k_0d$ varies from 0.1 to 1000. Therefore, the seismic waves contain both low and high frequency components in the scattering sense. Actually seismic waves with a center frequency greater than 1 Hz are generally regarded as high frequency waves. Thus the bifurcation phenomenon has significant implications for seismic wave explanation and imaging. As show in the previous work, it provides a new explanation to the observed Pn, Pg, Sn, Sg, P*, and S* phases of the seismic waves in the lithosphere, without invoking the layered model. In fact, the bifurcation of dispersion also happens to seismic waves in the mantle and the inner core, since they all exhibit certain degree of property fluctuation. In this sense, the occurrence of two propagation modes have strong potential in the explanation of multiple phases recorded in deep Earthquakes or teleseismograms, such as the mysterious high frequency precursors [69]. The magnitude and variation tendency of Q-factors are strongly dependent on the degree of material property fluctuation and the correlation length of the inhomogeneities, thus they can be used for inversion of the small-scale structure in the deep Earth. From this point of view, the model developed in this work and in the previous work [51] provide the theoretical foundation for statistical classification and modeling of the Earth.

Numerical simulation of coherent waves in heterogeneous media has drawn broad attentions from both the ultrasonics and seismological communities [56-57, 70-72], and meaningful results have been obtained. However, there are still several technical challenges for conducting more accurate simulations. Most of current simulations are limited to relatively low frequency regime, i.e., $k_0 d < 10$. To the author's best knowledge, the high frequency behavior ($k_0 d > 100$) has not been explored. In this case, each crystallite contains 20 to 50 wavelengths. To accurately capture the propagation characteristics, one should assign ten or more spatial steps in a single wavelength. Considering the fact that a statistically meaningful sample normally contains about 10000 crystallites, the computational cost is extremely expensive. Modern computational techniques like supercomputer, Graphic Processing Units, and parallel algorithms provide exciting opportunities for conducting such simulation. Another challenge is the generation of the microstructures. As the theory requires, the sample should be macroscopically homogeneous, the size distribution should be well controlled so that the exponential correlation function is well realized. The grain boundaries should be dealt with carefully so that the scattering characteristics are accurately reflected. Finally, the time evolution algorithms should be pure, free from artificial damping, numerical dispersion and excitation. To obtain the dispersion and attenuation that are statistically meaningful, a certain number of simulations should be conducted, normally on tens to a hundred samples. The most challenging point for conducting accurate experimental study is the manufacture of the sample. As pointed out by the pioneering researchers, there are always impurities exist in the sample, such as voids, twins, and other inclusions [30-31, 73-74]. The distribution of grain sizes is difficult to control accurately. Consequently, accurate and reliable experimental verification in the whole frequency range is still an open problem.

Finally, we explore possible future developments of current model. The current work only focuses on single phase polycrystals with equiaxed grains without preferred orientations. However, most practically important polycrystals exhibit more complicated microstructures due to different thermomechanical processing procedures. Thompson [10] classified these microstructures into four categories, 1) equiaxed grains without texture, 2) elongated grains without texture, 3) equiaxed grains with texture, 4) elongated grains with texture. For polycrystals with elongated grains and/or textures the scattering characteristics demonstrate obvious anisotropic behavior. Developing new multiple scattering theory which is capable of incorporating these complications has significant practical importance for quantitative characterization microstructures and imaging the small defects (like hard alpha-phase inclusions, microcracks etc.) in advanced engine alloys. Multi-phase alloy is another type of important polycrystalline materials. For single phase polycrystals, the material property fluctuation is solely due to elastic stiffness mismatch caused by random crystallographic orientations. The situation for multi-phase polycrystals are considerably more complicated since both mass density and elastic properties are random variables of spatial coordinates. As pointed out in [51], for this case the weak scattering theory gives unstable prediction for the propagation parameters. Nevertheless, the theory developed in this work is naturally suitable to study the ultrasonic scattering in multi-phased materials. Thus extending the current model for the case of multi-phased polycrystalline materials is one major task in future. The new model will enable the possibility of quantifying the volume ratio of the alpha phase to the beta phase in alpha+beta titanium alloys, which can be used to optimize the addition of alpha or beta stabilizers [15].

Accompanied with development of ultrasonic scattering model, different versions of ultrasonic grain noise models have also been proposed. The technical objective of grain noise model is two-folded, one is to properly remove the effects of grain noise and increase the capability of detecting and locating minor flaws in the buck materials, the other is to extract the microstructural information via backscattering coefficients or figure-of-merit (FOM). Rose [59-61] first obtained closed-form expressions of backscattering coefficients for polycrystalline materials with the use of independent scattering approximation (ISA) and the Born approximation. Later on, Han et. al. [15] extended the backscattering model to titanium alloys with duplex microstructures. Ghoshal and Turner [75-76] derived the ultrasonic backscatter signals in the context of diffuse backscatter measurements. They considered both multiple scattered mean and mean square response of the heterogeneous medium. Additionally, they introduced the Wigner transform techniques to describe the beam effects of a piston transducer. Turner and Weaver [77-78] also developed the radiative transfer theory to describe the diffusion of ultrasonic energy when the propagation distance is large and the ultrasound loss coherence. From the fundamental assumptions (ISA, the Born approximation, the weak anisotropy assumption, etc.) of the various grain noise models and radiative models, we see most of them are only applicable to polycrystals with weak property fluctuation, or only valid in the short-period of time when the multiple scattering has not fully developed. Based on the theoretical framework developed in this work, we can develop a most general radiative transfer theory and the corresponding grain noise model, which fully considers the multiple scattering effects, valid for strong scattering polycrystals, and puts no restrictions on the propagation time.

## VI. CONCLUSION

This work lays the new foundation of the multiple scattering theory for polycrystalline materials with strong grain anisotropy. The integral representation of the displacement and strain Green's function for polycrystalline materials is derived by using the homogeneous Green's functions. The singularity of the Green tensor is properly considered and the renormalized integral equations governing the new set of field variables are formulated. Feynman's diagram method is introduced to derive the renormalized Dyson's equation. The system of integral equations is solved using the Fourier transform technique. The dispersion equations for the coherent waves are obtained, from which the exact dispersion and attenuation curves for a large variety of polycrystalline materials in the whole frequency range are further calculated. The accuracy of the model is validated through comparison with experimental and numerical results. Comparison with experimental data shows the model is capable of predicting the average grain size with a relative error less than 10%. Comprehensive study on the dispersion and attenuation of most frequently used materials, including pure titanium, Ti64 and Ti6242 is carried out to show the applications of the new model in ultrasonic NDE. It is demonstrated the model is capable of quantitatively characterize the effects of phase composition, grain size and grain anisotropy on the velocity and attenuation of various titanium alloys. As for applications in seismology, the velocity and Q-factors of both hexagonal and cubic iron models of the Earth's inner core are calculated. Primary results show it is able to explain the observed phenomena like anormally slow transverse wave velocity and the P wave attenuation. This work establishes a new theoretical foundation for developing the next generation quantitative ultrasonic techniques for microstructure characterization in various polycrystalline materials, which is extreme importance for turbine jet engine manufacturing and inspection. The new model also demonstrates great potential for applications in seismic imaging and inversion of the material composition and structures in the Earth's inner core.

## Appendix Elastic stiffness of a general anisotropic material and its coordinate transform

In this appendix, we derive the relations between the components of the elastic stiffness tensor in different coordinate systems. Consider a crystallite (or a grain) of general anisotropy, it has 21 independent elastic constants, which can be expressed in the crystallographic coordinate system O-XYZ as:

$$C_{ijkl} = C_{pq} = \begin{bmatrix} C_{11} & C_{12} & C_{13} & C_{14} & C_{15} & C_{16} \\ C_{12} & C_{22} & C_{23} & C_{24} & C_{25} & C_{26} \\ C_{13} & C_{23} & C_{33} & C_{34} & C_{35} & C_{36} \\ C_{14} & C_{24} & C_{34} & C_{44} & C_{45} & C_{46} \\ C_{15} & C_{25} & C_{35} & C_{45} & C_{55} & C_{56} \\ C_{16} & C_{26} & C_{36} & C_{46} & C_{56} & C_{66} \end{bmatrix}. \tag{A1}$$

In this work, the elastic stiffness tensor in the crystallographic (local) coordinate system is represented by the capital letter **C**, the elastic stiffness tensor in the laboratory (global) coordinate system is represented by the lowercase letter **c**. The contracted indices *p* and *q* are related to the full tensorial indices *ij* and *kl* following the rule given in Tab. A1.

Table A1 Correspondence between contracted indices and the Cartesian tensor indices

| *ij, kl* | 11 | 22 | 33 | 23, 32 | 13, 31 | 12, 21 |
|---|---|---|---|---|---|---|
| *p, q* | 1 | 2 | 3 | 4 | 5 | 6 |

Polycrystalline materials are composed of a large number of grains. Each grain can be viewed as a piece of single crystal. Seven types of crystallographic symmetries have been identified, from high to low symmetries, they are cubic, hexagonal, tetragonal, trigonal, orthorhombic, monoclinic and triclinic crystals. To secure the generality of the theoretical derivation, we neglect the unique symmetry of different types of crystals, and treat them as generally anisotropic materials. In practical applications, we often need to use the elastic coefficients in a laboratory coordinate system O-xyz. The orientation of the crystallographic coordinate system relative to the laboratory coordinate system is described by three Euler angles ($\varphi, \theta, \psi$), as depicted in Fig. A1. The ranges of the three angles are:

$$0 \leq \varphi < 2\pi, \quad 0 \leq \theta \leq \pi, \quad 0 \leq \psi < 2\pi. \tag{A2}$$

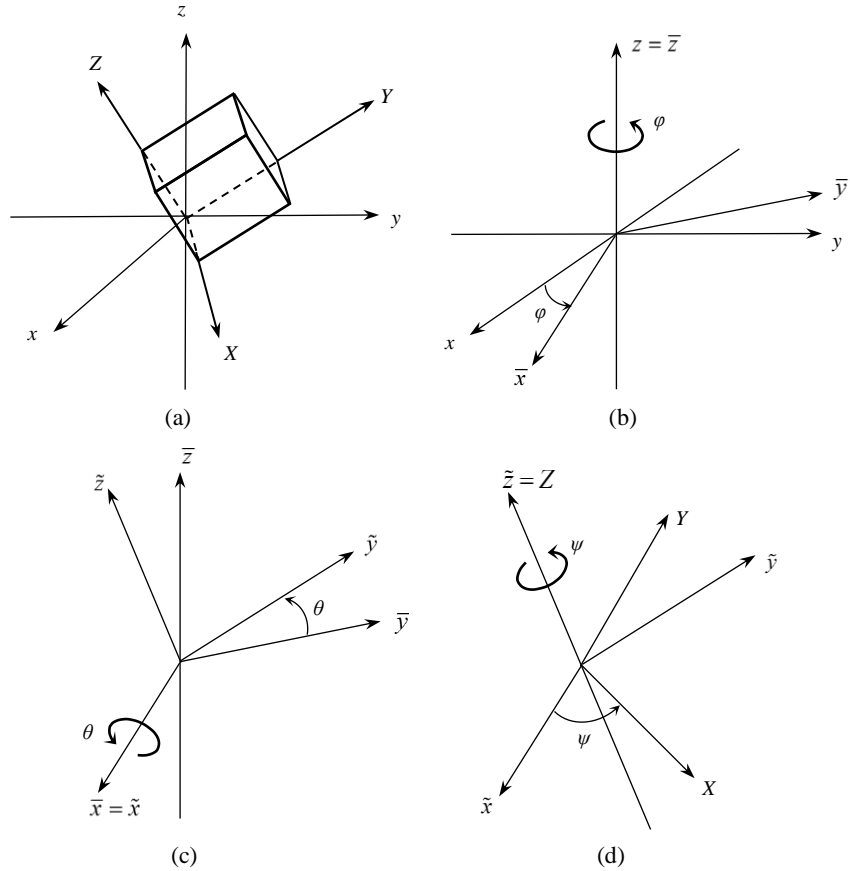

(a)　(b)　(c)　(d)

Figure A1　Definition of the Euler angles

The crystallographic coordinate system attached on each grain is denoted by O-XYZ, the coordinate basis corresponding to the X, Y and Z axes are denoted by $\mathbf{E}_1$, $\mathbf{E}_2$ and $\mathbf{E}_3$, respectively. The global coordinate system is denoted by O-xyz, its coordinate basis are $\mathbf{e}_1$, $\mathbf{e}_2$, $\mathbf{e}_3$. According to Euler's convention, the global coordinate frame can be transferred to the crystallographic coordinate frame through three successive rotations: the first step is rotating the O-xyz system about the z axis by an angle $\varphi$, we get another coordinate system O-$\bar{x}\bar{y}\bar{z}$, the corresponding coordinate basis are denoted by $\bar{\mathbf{e}}_1$, $\bar{\mathbf{e}}_2$, and $\bar{\mathbf{e}}_3$, the second step is rotating the O-

$\overline{x}\overline{y}\overline{z}$ system around the axis $\overline{x}$ by an angle $\theta$, we get a new coordinate system O-$\tilde{x}\tilde{y}\tilde{z}$, the coordinate basis are denoted by $\tilde{\mathbf{e}}_1$, $\tilde{\mathbf{e}}_2$ and $\tilde{\mathbf{e}}_3$, the third step is rotating O-$\tilde{x}\tilde{y}\tilde{z}$ around the $\tilde{z}$ axis by $\psi$, and then arrive at the crystallographic coordinate system O-XYZ. The coordinate basis has the following relations:

$$\begin{bmatrix}\overline{\mathbf{e}}_1\\\overline{\mathbf{e}}_2\\\overline{\mathbf{e}}_3\end{bmatrix}=\begin{bmatrix}\cos(\varphi) & \sin(\varphi) & 0\\-\sin(\varphi) & \cos(\varphi) & 0\\0 & 0 & 1\end{bmatrix}\begin{bmatrix}\mathbf{e}_1\\\mathbf{e}_2\\\mathbf{e}_3\end{bmatrix},\quad [\overline{\mathbf{e}}]=\mathbf{R}(\varphi)[\mathbf{e}]\,,\tag{A3}$$

$$\begin{bmatrix}\tilde{\mathbf{e}}_1\\\tilde{\mathbf{e}}_2\\\tilde{\mathbf{e}}_3\end{bmatrix}=\begin{bmatrix}1 & 0 & 0\\0 & \cos(\theta) & \sin(\theta)\\0 & -\sin(\theta) & \cos(\theta)\end{bmatrix}\begin{bmatrix}\overline{\mathbf{e}}_1\\\overline{\mathbf{e}}_2\\\overline{\mathbf{e}}_3\end{bmatrix},\quad [\tilde{\mathbf{e}}]=\mathbf{R}(\theta)[\overline{\mathbf{e}}]\,,\tag{A4}$$

$$\begin{bmatrix}\mathbf{E}_I\\\mathbf{E}_{II}\\\mathbf{E}_{III}\end{bmatrix}=\begin{bmatrix}\cos(\psi) & \sin(\psi) & 0\\-\sin(\psi) & \cos(\psi) & 0\\0 & 0 & 1\end{bmatrix}\begin{bmatrix}\tilde{\mathbf{e}}_1\\\tilde{\mathbf{e}}_2\\\tilde{\mathbf{e}}_3\end{bmatrix},\quad [\mathbf{E}]=\mathbf{R}(\psi)[\tilde{\mathbf{e}}]\,,\tag{A5}$$

By consecutive substitution from (A3) into (A4) and then into (A5), we obtain the transformation relation between the two sets of bases:

$$[\mathbf{E}]=\mathbf{Q}[\mathbf{e}]\,,\quad [\mathbf{e}]=\mathbf{Q}^T[\mathbf{E}]\,,\tag{A6}$$

where

$$\mathbf{Q}=\mathbf{R}(\psi)\mathbf{R}(\theta)\mathbf{R}(\varphi)\,,\tag{A7}$$

$$\mathbf{Q}=\begin{bmatrix}q_{11} & q_{12} & q_{13}\\q_{21} & q_{22} & q_{23}\\q_{31} & q_{32} & q_{33}\end{bmatrix}=\begin{bmatrix}\cos(\mathbf{e}_1,\mathbf{E}_1) & \cos(\mathbf{e}_2,\mathbf{E}_1) & \cos(\mathbf{e}_3,\mathbf{E}_1)\\\cos(\mathbf{e}_1,\mathbf{E}_2) & \cos(\mathbf{e}_2,\mathbf{E}_2) & \cos(\mathbf{e}_3,\mathbf{E}_2)\\\cos(\mathbf{e}_1,\mathbf{E}_3) & \cos(\mathbf{e}_2,\mathbf{E}_3) & \cos(\mathbf{e}_3,\mathbf{E}_3)\end{bmatrix}$$
$$=\begin{bmatrix}\cos(\varphi)\cos(\psi)-\sin(\varphi)\sin(\psi)\cos(\theta) & \sin(\varphi)\cos(\psi)+\cos(\varphi)\sin(\psi)\cos(\theta) & \sin(\psi)\sin(\theta)\\-\cos(\varphi)\sin(\psi)-\sin(\varphi)\cos(\psi)\cos(\theta) & -\sin(\varphi)\sin(\psi)+\cos(\varphi)\cos(\psi)\cos(\theta) & \cos(\psi)\sin(\theta)\\\sin(\varphi)\sin(\theta) & -\cos(\varphi)\sin(\theta) & \cos(\theta)\end{bmatrix}\tag{A8}$$

When the orthogonal transformation matrix $\mathbf{Q}$ is given, we can also solve for the Euler's angles. This procedure is particularly useful in analyzing texture evolution of polycrystalline materials.

The components of the elastic stiffness tensor in the crystallographic coordinate system and those in the global coordinate system can be transformed between each other as follows:

$$\mathbf{c}=\mathbf{N}^T\mathbf{C}\mathbf{N}\,,\quad \mathbf{C}=\mathbf{M}\mathbf{c}\mathbf{M}^T\,.\tag{A9}$$

where

$$\mathbf{M}=\begin{bmatrix}q_{11}^2 & q_{12}^2 & q_{13}^2 & 2q_{12}q_{13} & 2q_{11}q_{13} & 2q_{12}q_{11}\\q_{21}^2 & q_{22}^2 & q_{23}^2 & 2q_{23}q_{22} & 2q_{21}q_{23} & 2q_{22}q_{21}\\q_{31}^2 & q_{32}^2 & q_{33}^2 & 2q_{32}q_{33} & 2q_{31}q_{33} & 2q_{31}q_{32}\\q_{21}q_{31} & q_{32}q_{22} & q_{23}q_{33} & q_{22}q_{33}+q_{32}q_{23} & q_{21}q_{33}+q_{23}q_{31} & q_{21}q_{32}+q_{22}q_{31}\\q_{11}q_{31} & q_{12}q_{32} & q_{13}q_{33} & q_{12}q_{33}+q_{13}q_{32} & q_{11}q_{33}+q_{31}q_{13} & q_{11}q_{32}+q_{31}q_{12}\\q_{11}q_{21} & q_{12}q_{22} & q_{13}q_{23} & q_{12}q_{23}+q_{13}q_{22} & q_{11}q_{23}+q_{13}q_{21} & q_{11}q_{22}+q_{21}q_{12}\end{bmatrix}.\tag{A10}$$

$$\mathbf{N}=\begin{bmatrix}q_{11}^2 & q_{12}^2 & q_{13}^2 & q_{13}q_{12} & q_{11}q_{13} & q_{12}q_{11}\\q_{21}^2 & q_{22}^2 & q_{23}^2 & q_{23}q_{22} & q_{21}q_{23} & q_{22}q_{21}\\q_{31}^2 & q_{32}^2 & q_{33}^2 & q_{33}q_{32} & q_{31}q_{33} & q_{32}q_{31}\\2q_{31}q_{21} & 2q_{32}q_{22} & 2q_{33}q_{23} & q_{32}q_{23}+q_{33}q_{22} & q_{33}q_{21}+q_{31}q_{23} & q_{31}q_{22}+q_{21}q_{32}\\2q_{11}q_{31} & 2q_{12}q_{32} & 2q_{13}q_{33} & q_{12}q_{33}+q_{13}q_{32} & q_{13}q_{31}+q_{11}q_{33} & q_{11}q_{32}+q_{12}q_{31}\\2q_{21}q_{11} & 2q_{22}q_{12} & 2q_{23}q_{13} & q_{22}q_{13}+q_{12}q_{23} & q_{23}q_{11}+q_{21}q_{13} & q_{11}q_{22}+q_{21}q_{12}\end{bmatrix},\tag{A11}$$

It can be shown that:

$$\mathbf{M}^{-1}=\mathbf{N}^T\,,\quad \mathbf{M}^T=\mathbf{N}^{-1}\,,\tag{A12}$$

From Eq. (A9) we can see the elastic coefficients of each grain in the global system is a function of its orientation:

$$c_{ijkl} = c_{ijkl}(\varphi, \theta, \psi). \tag{A13}$$

The fluctuation of elastic stiffness with respect to the homogeneous reference medium is:

$$\delta c_{ijkl}(\varphi, \theta, \psi) = c_{ijkl}(\varphi, \theta, \psi) - c_{ijkl}^{(0)}, \tag{A14}$$

where $c_{ijkl}^{(0)}$ is the elastic stiffness of the reference medium:

$$c_{ijkl}^{(0)} = \lambda \delta_{ij}\delta_{kl} + \mu(\delta_{ik}\delta_{jl} + \delta_{il}\delta_{jk}). \tag{A15}$$

Substituting from Eq. (A14) into Eq. (25) gives $\Xi(\mathbf{x})$, which is also a function of the Euler angles. The ensemble average of the elastic quantity $\Xi(\mathbf{x})$ is taken over all the possible orientations, i.e., over the SO(3) group.